\pdfoutput=1 
\documentclass[twocolumn]{aastex62}
\usepackage{lipsum}% http://ctan.org/pkg/lipsum
\usepackage[flushleft]{threeparttable}
\usepackage{scrextend}
\usepackage{graphicx}
\usepackage{amsmath}
\usepackage{ulem}
\usepackage[utf8]{inputenc}
\usepackage{amssymb}
\usepackage[normalem]{ulem}
\received{July 16, 2018}
\revised{December 05, 2018}
\accepted{March 4, 2019}
\submitjournal{ApJ}

\begin{document}

\title{Evolution and Spectral Response of a Steam Atmosphere for Early Earth with a coupled climate-interior model}

\correspondingauthor{N. Katyal}
\email{nisha.katyal@dlr.de}

\author[0000-0002-0786-7307]{Nisha Katyal}
\affil{Zentrum fur Astronomie und Astrophysik, Technische Universität, Hardenbergstr., 10623 Berlin, Germany}
\affil{Institut für Planetenforschung, Deutsches Zentrum für Luft- und Raumfahrt, Rutherfordstr. 2, 12489 Berlin Germany}

\author{Athanasia Nikolaou}
\affil{Zentrum fur Astronomie und Astrophysik, Technische Universität, Hardenbergstr., 10623 Berlin, Germany}
\affil{Institut für Planetenforschung, Deutsches Zentrum für Luft- und Raumfahrt, Rutherfordstr. 2, 12489 Berlin, Germany}

\author{Mareike Godolt}
\affil{Zentrum fur Astronomie und Astrophysik, Technische Universität, Hardenbergstr., 10623 Berlin, Germany}

\author{John Lee Grenfell}
\affil{Institut für Planetenforschung, Deutsches Zentrum für Luft- und Raumfahrt, Rutherfordstr. 2, 12489 Berlin, Germany}

\author{Nicola Tosi}
\affil{Zentrum fur Astronomie und Astrophysik, Technische Universität, Hardenbergstr., 10623 Berlin, Germany}
\affil{Institut für Planetenforschung, Deutsches Zentrum für Luft- und Raumfahrt, Rutherfordstr. 2, 12489 Berlin, Germany}

\author{Franz Schreier}
\affil{Institut für Methodik der Fernerkundung
, Deutsches Zentrum für Luft- und Raumfahrt, 82234 Oberpfaffenhofen, Germany}

\author{Heike Rauer}
\affil{Department of Planetary Sciences, Institute of Geosciences, Freie Universität, Malteserstr. 74-100, 12249 Berlin, Germany}
\affil{Institut für Planetenforschung, Deutsches Zentrum für Luft- und Raumfahrt, Rutherfordstr. 2, 12489 Berlin, Germany}
\affil{Zentrum fur Astronomie und Astrophysik, Technische Universität, Hardenbergstr., 10623 Berlin, Germany}

\begin{abstract}
The evolution of Earth's early atmosphere and the emergence of habitable conditions on our planet are intricately coupled with the development and duration of the magma ocean phase during the early Hadean period (4 to 4.5 Ga). In this paper, we deal with the evolution of the steam atmosphere during the magma ocean period. We obtain the outgoing longwave radiation using a line-by-line radiative transfer code GARLIC. Our study suggests that an atmosphere consisting of pure H$_{2}$O, built as a result of outgassing extends the magma ocean lifetime to several million years. The thermal emission as a function of solidification timescale of magma ocean is shown. We study the effect of thermal dissociation of H$_{2}$O at higher temperatures by applying atmospheric chemical equilibrium which results in the formation of H$_{2}$ and O$_{2}$ during the early phase of the magma ocean. A 1-6\% reduction in the OLR is seen. We also obtain the effective height of the atmosphere by calculating the transmission spectra for the whole duration of the magma ocean. An atmosphere of depth ~100 km is seen for pure water atmospheres. The effect of thermal dissociation on the effective height of the atmosphere is also shown. Due to the difference in the absorption behavior at different altitudes, the spectral features of H$_{2}$ and O$_{2}$ are seen at different altitudes of the atmosphere. Therefore, these species along with H$_{2}$O have a significant contribution to the transmission spectra and could be useful for placing observational constraints upon magma ocean exoplanets.

\end{abstract}


\section{Introduction} \label{sec:intro}

The early atmospheric evolution of Earth and other terrestrial planets is likely to be characterized by the presence of one or several Magma Oceans (MOs) with a hot and largely molten mantle lying above the core \citep{Elkins2012}. A magma ocean consists of molten silicates undergoing turbulent convection and is a natural outcome of core formation and giant energetic impacts during planetary accretion \citep{Bonati2019} or via comets and meteorites \citep*{Abe1985,Canup2004}. 
%{\color{red} The atmospheres of Earth-like planets are indicative of the fact that impacts have occurred \citep{Lupu2014}. } 

During the accretion of planetesimals and following giant impacts, a transient atmosphere due to impact devolatilization was present above the magma ocean, but the planet did not acquire its initial water content through ingassing from such an atmosphere. Recent models of ingassing of nebular hydrogen into the magma ocean show that only 1\% of water entered the planet in this way \citep{Wu18}. The current general consensus is that the Earth acquired its water inventory largely via accretion of chondritic materials \citep{review17}. Also, experiments of metal-silicate partitioning require the presence of a hydrous magma ocean to account for the characteristics of certain trace-elements of the Earth \citep[e.g.][]{right99}. At any rate, when the  Earth’s magma ocean started to solidify, it already had a significant amount of water to be outgassed, no matter how it was acquired.

The chemical composition of the atmosphere built as a result of outgassing from the interior depends upon the compositions of different meteoritic classes materials e.g., carbonaceous, ordinary, or enstatite chondrites as discussed by \citet*{SF2007,SF2010}. Furthermore, it has been argued that a mantle rich in hydrogen would give rise to a reduced atmosphere mainly consisting of CO, CH$_{4}$, NH$_{3}$ and H$_{2}$, whereas a mantle more abundant in oxygen would give rise to an oxidized atmosphere mainly consisting of CO$_2$ and H$_{2}$O \citep*{Zahnle10,Gail14}. Recently, the oxygen fugacities for the different meteoritic materials (both reduced and oxidized) have been explored in greater details by \citet{SF2017}. In the present paper, we investigate the atmospheric evolution by assuming an oxidized mantle whereby the important factor remains the planetary outgassing. 

Atmospheric 1D-column models focusing on the ``greenhouse effect" with surface temperatures exceeding the critical temperature of water (647 K) have been employed by \citet{Kasting88} and \citet{Naka92} that obtained an Outgoing Longwave Radiation (OLR) limit of 280-300 W m$^{-2}$. Therefore, water photolysis and H escape are both significant processes that impact the atmospheric composition and mass as shown by \citet{Schaefer16}. H escape may lead to the build up of O$_{2}$. The planet GJ1132b studied by \citet{Schaefer16} could be trapped in a long term MO, so water photolysis followed by H-escape could be important processes. \citet{Gold13}  studied the transition to runaway greenhouse and obtained a radiation limit of 282 W m$^{-2}$ for modern Earth. This limit, however, is obtained for $T_{\rm s} < 1800$ K and breaks down when the surface becomes hot enough to radiate in the visible wavelength regime and then the OLR is seen to rise.

Usually in these models, the atmosphere is assumed to be in global energy balance, i.e. OLR or the emitted radiation by the planet is equal to the absorbed shortwave incoming radiation from the host star. This condition, however, might not be satisfied for cases with a deep MO lying below a massive steam atmosphere formed as a result of an additional heat flux from the interior. Such scenarios require more detailed investigations \citep[e.g.][]{Hamano13}. 
%More recently, a 1D radiative-convective atmospheric model for surface temperature up to the critical temperature of water has been employed to investigate the effect of outgassed CO$_2$ and H$_2$O upon the boundaries of the Habitable Zone (HZ) of a ``stagnant-lid Earth" \citep{Tosi17}.

In the case of coupled atmospheric-interior models, the evolution of the atmosphere and thermal cooling of the MO are invariably linked to each other and are influenced by several assumptions in the numerical modeling. First of all, the assumption of a grey or non-grey radiative transfer approach in the atmospheric models. For example, various studies linking the atmosphere-interior model have used a \emph{grey} radiative transfer approach \citep*{Matsui1986,Elkins08,Lebrun13}. While, \citet*{Hamano15,Schaefer16} have used a non-grey approach.
%These models assumed  as the main radiative species and the atmospheric models used in these studies are inspired by the study of \citet{Kasting88}. 
Secondly, the composition of the atmosphere. A radiative-convective 1D-atmospheric model assuming pure water composition, and based on the \emph{grey} approach of \citet{Naka92} has been coupled with an interior model by \citet{Hamano13}. On the other hand, both H$_2$O and CO$_{2}$ have been assumed as the main radiative species by \citet*{Matsui1986,Elkins08,Lebrun13}. It has been shown by \citet{Elkins11} that a MO with as low as 0.1 wt \% (1000 ppm) of water can potentially outgas hundreds of bars of water. Moreover, the formation of water oceans on rocky planets (e.g., due to the collapse of a steam atmosphere) suggests that steam (H$_2$O) is the major volatile reservoir for most planetary surfaces and their mantles. Therefore, in this paper, we use only water as the constituent of the atmosphere. 

Finally, various parameterizations in the interior model also affect the thermal cooling of the MO. For example, the cooling of the MO might also get influenced by factors such as the initial volatile content in the molten mantle \citep{Sal17}. The influence of initial volatile water content (varied from  0.1 to up to 1000 mass of Earth ocean $M_\text{EO}$) upon the solidification timescale of MO has been explored by \citet{Schaefer16}, with a focus on the atmospheric escape of H and O. Other factors affecting the delay of the MO are the viscosity of the fluid and the depth of the magma ocean \citep*{Solom2007,Lebrun13}, as well as assumptions in the melting and solubility curves which are explored with more details in \citet{Nasia18}. 

During the MO solidification phase, the growth and evolution of an atmosphere also affects the planet's thermal spectra significantly. In order to study the spectral evolution of hot molten planets with up to 50 Earth Oceans ($M_\text{EO}$) of water, a \emph{non-grey} radiative transfer model was developed by \citet{Hamano15}. A planet with bulk initial water content exceeding 1 wt \% would outgas large volumes of volatile into the atmosphere. Their work describes the growth of a massive steam atmosphere during the MO solidification phase. And their calculations have indicated that the blanketing effect of a``water-dominated" atmosphere prevents the thermal radiation from escaping the planet, and prolongs the MO solidification timescale to $\sim$ 4 Myr for type I planets and 100 Myr for type II planets \citep{Hamano13}. The planets, designated as type I and type II by  are located at a critical distance  of 1 AU and 0.7 AU from the host star, respectively. 

Very recently, \citet{Bonati2019} have discussed the possible direct detection of magma ocean planets in nearby young stellar associations with future facilities for example, E-ELT\footnote{European-Extremely Large Telescope} and LIFE\footnote{Large Interferometer for Exoplanets}. As shown by them, the young and close stellar objects (Table 1 of \cite{Bonati2019}) with highest frequencies of giant impacts (e.g. $\beta$ Pictoris having an age of 23 Myr) leading to the formation of MOs have a higher probability for detection of magma oceans. For a steam atmosphere, only a few planets with a flux density of the order $10^{-2}-10^{-3} (\mu \rm\,Jy)$ would be detectable in the $\beta$ Pictoris stellar association with an observation time of 50 hrs with E-ELT through direct imaging \citep{Bonati2019}. The PLATO\footnote{Planetary Transits and Oscillations of stars} mission, on the other hand will provide accurate ages to a precision of 10\% and radii to a precision of 2\% for a large sample of planetary systems \citep{Rauer2014} which will lead to better characterization of young magma ocean planets. Additionally, accurate estimation of brightness temperature of a planet and the contrast ratio between planet and star can determine whether the atmospheric signatures are detectable or not \citep{Lupu2014,Hamano15,Marcq17}. Such studies motivate the investigation of exoplanetary atmospheres through the emission and transmission spectra which can provide information on the planet's subsequent evolution to habitable conditions.

%At high surface temperatures such as those suggested for the MO period, effects such as the atmospheric escape of H due to the high XUV stellar flux, photo-dissociation and the thermal dissociation of H$_2$O to form additional species becomes important. In the latter case, H$_2$O might be present only in the cooler upper layers of the atmosphere while lower layers (close to the surface) would mainly consist of other H and O bearing species e.g., H$_2$ and O$_2$ due to slow diffusion rates. 
During the Hadean period, effects such as photolysis of water molecules
and the atmospheric escape of H due to the high stellar irradiation,
and thermal dissociation of H$_{2}$O due to the high surface temperatures
will lead to the formation of additional species. In the latter case, the atmosphere in the lower layers close to the surface would mainly consist of other
 H and O bearing species e.g., H$_{2}$ and O$_{2}$. The build-up of abiotic O$_{2}$ under such conditions has also been discussed in detail by \citet{Kasting1995} and \citet{Schaefer16}. In this paper, we investigate the possible abiotic build-up of O$_{2}$ and other O bearing species due to thermal dissociation of water.  We also show the contribution of these newly formed species to the transmission spectra as a function of wavelength.

%This facilitates us to study the atmospheric evolution by using a \emph{non-grey}  H$_2$O gas in our radiative transfer calculations enabling us to evaluate the thermal spectra of hot planets with magma oceans.

\begin{table*}[!ht]
\small
\label{table1}
\centering
\caption{A comparison table for the recent atmospheric model studies with coupling to an interior model. }
\begin{tabular}{|l |c|c |c|c|c|c|c|c|}
\hline\hline
Reference & Composition  & Atm. Type & Atm. Pressure & Rad. Transfer & OLR limit  & Rayleigh & Clouds & Addition\\
         &              &               &  (bar) &  method     &  (Wm$^{-2}$)&   scattering      &        &        \\ 
\hline
\citet{Marcq12}           & H$_{2}$O-CO$_{2}$ &Non-grey & 270 & correlated-$k$& 164 & Yes& Yes & None\\
\citet{Lebrun13}          & H$_{2}$O- CO$_{2}$&grey & -       & correlated-$k$& No & -& - & None \\
%%                       &                   &         & (SMART)       &     &    &    &         \\
%%Koparappu et al. (2013)& H$_{2}$O-CO$_{2}$ & Non-grey&correlated-$k$& 291  & No & No & No\\
\citet{Hamano13} &   H$_{2}$O          & Grey & 10-1250 & correlated-$k$              & 280 & Yes& No & None\\
\citet{Hamano15} &   H$_{2}$O         & Non-grey & 5 $\times$ 10$^{-4}$-5000 &correlated-$k$ & 280 & Yes& No & Atm. escape\\
\citet{Schaefer16}& H$_{2}$O         & Non-grey &270 &line-by-line   & --  &--  & No & Atm. escape\\
                      &                      &      & & (SMART)       &     &    &    &         \\
\citet{Marcq17} &   H$_{2}$O-CO$_{2}$ & Non-grey &270 &correlated-$k$ & 280 & Yes & Yes & None\\
This study (2018) &  H$_{2}$O            & {\bf Non-grey} & {\bf 4-300} &line-by-line  & 282 & No & No &  {\bf Thermal}  \\
                  &                      &          & &{\bf (GARLIC)}      &     &    &    &  {\bf dissociation}       \\
		  &                      &          & &      &         &    &   &\& effective       \\
		  &                      &          & &      &         &    &  &atm. height       \\
%&  &  & & &  & \\
%&  &  & &  & & \\
 \hline
\end{tabular}
\end{table*}

Table 1 provides a comparison of different coupled interior-atmospheric models from the literature along with the methods used for radiative transfer calculations in each of them.

With the growing need for high resolution infrared (IR) and microwave spectra to compare with the observations of the planets, Line-by-Line (lbl) modeling of atmospheric radiative transfer is essential. Therefore, in this paper, we utilize the lbl radiative transfer model \software{GARLIC (Schreier et al. 2014)} (discussed in detail in Sect.~\ref{lbl}) to calculate the emission from a steam atmosphere, which is formed as a result of outgassing from the Earth's interior. The main motivation of our study is to investigate the impact of a time-dependent outgassing of volatiles on the thermal and chemical evolution of a steam atmosphere during the magma ocean phase of the early Earth. Our calculation of the atmospheric radiative flux for H$_{2}$O without the radiative effects of H$_{2}$O speciation into H$_2$ and O$_{2}$ is used by the companion paper to assess the changes in the surface temperature and to obtain the solidification timescale of the magma ocean. Also, the companion paper discusses the effects of several parameters that affect the MO cooling process. The evolution of surface temperature as a function of time is presented in the companion paper and this paper as well. Finally, we  study the effects of thermal dissociation of water on the OLR and the transmission spectra.

A brief outline of the paper is as follows. Section~\ref{atmos} describes the numerical models and the framework. Section~\ref{thermal} shows results of the outgoing longwave radiation (OLR) as a function of a prescribed [$T$,$p$] grid, which are an input to the interior model. The results of both models are combined together and presented in Section~\ref{couple}. In Section~\ref{trans}, the results of the wavelength-dependent effective height  of the atmosphere for the whole duration of magma ocean are presented. Section~\ref{cea} discusses the effects of thermal dissociation of water on the outgoing longwave radiation and the effective height of the atmosphere. Finally, we provide a discussion in Section~\ref{summ} followed by summary and conclusions in Section~\ref{Conc1}.

\section{Numerical Models and Framework} \label{atmos}

\subsection{Interior Model \label{int}}
Thermal cooling of the planetary interior and the outgassing of volatile species (H$_{2}$O here) is obtained with the use of a 1-D parameterized interior model. The model consists of a spherically symmetric mantle divided into layers of 1 km thickness assumed for a global MO stage for an Earth-sized planet. To ensure global mantle melting, we assume an initial mantle potential temperature of 4000 K. However, this can vary with different mantle melting curves. For the melting curves we use, the mantle is molten by assuming a potential temperature ($T_p$) of 3000 K and above. We start our calculations at $T_{\text p} = 4000$ K to be consistent with former studies \citep*{Lebrun13,Schaefer16} that calculate solidification timescales of the magma ocean. The vaporization of minerals at such high temperatures leading to the formation of ``silicate clouds'' is not included since the phase during which they were present was relatively very short $\sim 1000$ yrs \citep{Zahnle05}. However, the vaporization of silicate minerals is important for estimating the atmospheric type in terms of e.g. temperature and the relative amounts of vapourised metal and mineral species as shown by \citet{Miguel11}, for example, via cloud formation. However, more extensive laboratory data for the equilibrium composition of volatilised silicate minerals in hot steam atmospheres would be desirable. 

The hot magma ocean emits a heat flux, $F_\text{MAG}$ from the interior via the planetary surface due to turbulent convection. The temperature profile, i.e. the variation of surface temperature ($T_{\text s}$) with pressure ($p$) in the mantle is assumed to be adiabatic. The melt fraction $\phi(r)$ in the interior is calculated using the equation \citep{Solom2007}:

\begin{equation}
\phi(r)=\frac{T_{\text s}(r)-T_\text{sol}(r)}{T_\text{liq}(r)-T_\text{sol}(r)}, \label{e01}
\end{equation}
where $T_\text{sol}$ and $T_\text{liq}$ are the solidus and liquidus temperatures, respectively. The critical melt fraction is taken to be $\phi_{c}=0.4$ \citep{Lebrun13,Schaefer16} and divides the mantle into two zones. If $\phi > \phi_{c}$, the melt has a liquid-like behavior and if $\phi < \phi_{c}$, melt exhibits a solid-like behavior indicating the end of the MO phase and the simultaneous volatile outgassing from it. The critical value of the melt fraction separates the liquid-like magma from the solid-like mantle phase and the limit between the two zones is known as the rheology front (RF).  The mantle temperature corresponding to the rheology front at the surface denoted as $T_\text{RF,0}$ is a function of specific melting curves, explained in more details in the companion paper. 

The outgassing of water as a function of temperature is calculated using the 1D interior model assuming phase equilibrium with the melt and using the solubility curve of vapor for a given water concentration in the silicate melt \citep{Caroll94}. Comparing the interior temperature profile with the melting curves of the mantle, adopted by assuming a non-evolving KLB-1 peridotitic composition \citep{Zang,Herz,Hirsch,Fiquet}, yields the volume fraction of the mantle in the solid and the liquid phases \citep[e.g.][]{Nasia18}. The fractionation of the water vapour into the two respective reservoirs is then calculated, and the amount of water that is super-saturated in the melt is released into the atmosphere. Thus, the pressure of outgassed water vapour at each stage of the simulation is obtained. The magma ocean stage comes to an end when the potential temperature reaches $T_\text{RF,0}$ at the surface, i.e. 40 \% melt at the surface. For the melting curves employed in this study, this happens at $T_\text{RF,0} \sim 1650$ K.

By the use of the same 1D interior model, the convective heat flux out of the magma ocean is calculated. 

%\begin{equation}
%\frac{dT_{p}}{dt} \propto R_{p}^{2} F_{\rm MAG}. \label{e2}
%\end{equation}

%Here, F$_{\rm MAG}$ is a function of the potential temperature $T_{p}$ and the surface temperature $T_{s}$. 

%%The flux of the outgoing longwave radiation, F$_{\rm OLR}$ is evaluated on a grid of surface temperature and pressure [T,p] as described in detail in Sect.~\ref{lbl}. The method of bilinear interpolation is employed to evaluate the OLR for every $T_{s}$ that is intermediate to the chosen grid point. The solution for T$_{s}$ is numerically found by using an iterative method, similar to the one used in \citet{Lebrun13}. The fluxes are calculated at each iterative step and a new value for the T$_{s}$ is employed until the two fluxes converge to the same value. To satisfy the convergence criterion, a tolerance value for the difference of the two fluxes is set to 0.01 $\rm Wm^{-2}$. More details will be provided in Sect.~\ref{couple}.

\begin{figure}
\centering
\includegraphics[scale=0.6]{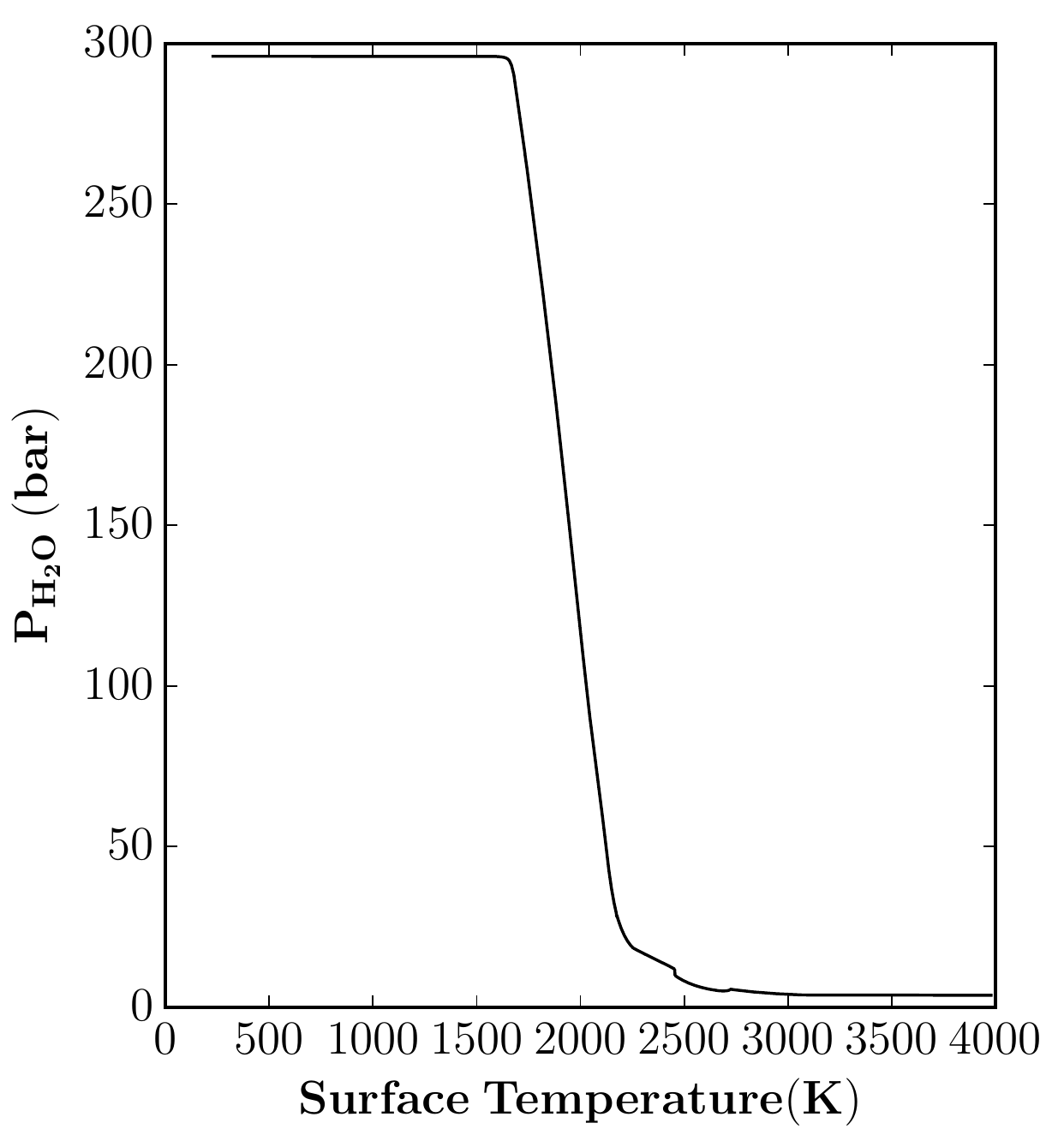}
%\plotone{figures/Ts_profiles_new}
\caption{Variation of surface temperature with the outgassed water pressure obtained from \citet{Nasia18} using an interior model coupled with a grey atmospheric model. The thermal emission is obtained for a reasonable range of surface temperature-pressure combinations [$T$,$p$] from this data set using the LBL model. \label{tp}}
\end{figure}

The initial water inventory chosen by us is slightly more than 1 Earth Ocean (EO$=1.39 \times 10^{21} $kg). It is equivalent to 550 ppm or 400 bar. The motivation for this was to investigate the extent by which these starting conditions could reproduce Earth's ocean reservoir of 300 bar. Using this water inventory, \citet{Nasia18} have obtained the outgassing of water from the mantle using an interior model coupled with a grey atmospheric model, and calculated the variation of surface temperature with the outgassed pressure, denoted as [$T$,$p$] henceforth and shown in Fig.~\ref{tp}. As an assumption, we prescribe their model output as the surface temperature and pressure for our atmosphere model. The range of outgassed pressure during the MO solidification corresponds to $P_{\rm H_2 O}$ $\sim$ [4, 300] bar (see Fig.~\ref{tp}) which is also the range of the input chosen for the calculation of atmospheric structure and flux of the outgoing longwave radiation, $F_{\rm OLR}$ as explained below.

%\subsection{Line-by-line (LBL) Atmospheric Model \label{lbl}}
\subsection{Atmospheric structure \label{str}}

In this section, we derive the atmospheric structure for the prescribed surface temperature and pressure as shown in Fig.~\ref{tp}. The atmosphere is assumed to be consisting of only H$_2$O and is divided into $N=66$ atmospheric layers equally spaced on a logarithmic scale of pressure and calculated by fixing the pressure at the top of the atmosphere to be 1 Pa. The temperature structure of the atmosphere is obtained by assuming a dry adiabatic lapse rate for the atmospheric layers whose temperature is above the critical temperature of water (647 K).  Since our calculations begin with a low pressure (4 bar) of water at higher temperatures, therefore, an assumption of the ideal gas equation for the calculation of dry adiabatic lapse is reasonable for our purpose. However, we additionally consider the T-dependence of $C_{p}$ over the whole range of temperatures and pressures. The dry adiabatic lapse rate is calculated from the surface to the top by integrating the equation: $d \ln(T)/d\ln(P) = R/C_{p}(T)$, where $R$ is the gas constant and $C_{p}(T)$ is the specific heat capacity of water with a temperature dependence, which is obtained using the following equation taken from \citet{Wagner02} (and also mentioned by \citet{Hamano13}):

\begin{equation}
\frac{C_{p}(T)}{R} = 1+n_{3}+ \sum_{i=4}^{8}n_i \frac{(\gamma_i \tau)^{2}\exp(-\gamma_i \tau)}{[1-\exp(-\gamma_i \tau)]^{2}}, \label{e7}
\end{equation}
where $\tau=T_{\text c}/T$ and $T_{\text c}$ is the critical temperature (647 K). The value of coefficients $n_{i}$ and $\gamma_{i}$ are taken from Table 6.1 of \citet{Wagner02}. 

The remaining procedure for calculating the atmospheric structure is the same as the \citet{Gold13}. The tropopause temperature $T_{\rm 0}$ for high surface temperature runs results from the moist adiabatic lapse rate when the pressure at the top of the atmosphere reaches 1 Pa. $T_{\rm 0}$ for our runs ranges from 352 K to 204 K (for decreasing surface temperatures) and it reaches 204 K at $T_{\text s}=2900 \rm \, K$. For departures from ideal gas behavior, a self-consistent derivation of moist adiabat requires the use of a non-ideal gas equation at high densities \citep{Pierre13}. 
%\sout{As described before, for the calculation of atmospheric structure, a dry adiabatic lapse rate is assumed for high temperatures until it intersects with the saturation water vapour curve wherein a moist adiabatic lapse rate is assumed for the rest of the atmospheric layers. The tropopause temperature is set to be 200 K. The scale height of the atmosphere is calculated by using the equation: $H_{s,i}=k_{B}\bar T_{i}/\mu g$; where the mean temperature $\bar T$ is given by $\bar T = (T_i + T_{i+1})/2$; $i=0,..N-1$ is the layer number; $\mu$ is the mean molecular weight of H$_2$O ($\mu=18$ amu), $k_{B}$ is the Boltzmann constant and $g=9.8$ m/s$^{2}$. The height of the atmosphere or the altitude (in km) is then calculated based on log pressure points using
%\begin{equation}
%Z = \sum_i^{N-1} H_{s} \log(\bar p),
%\end{equation}
%where $\bar p= (p_i/ p_{i+1})$ is defined to be the pressure between the two layers.}

\begin{figure}
\centering
\includegraphics[scale=0.8]{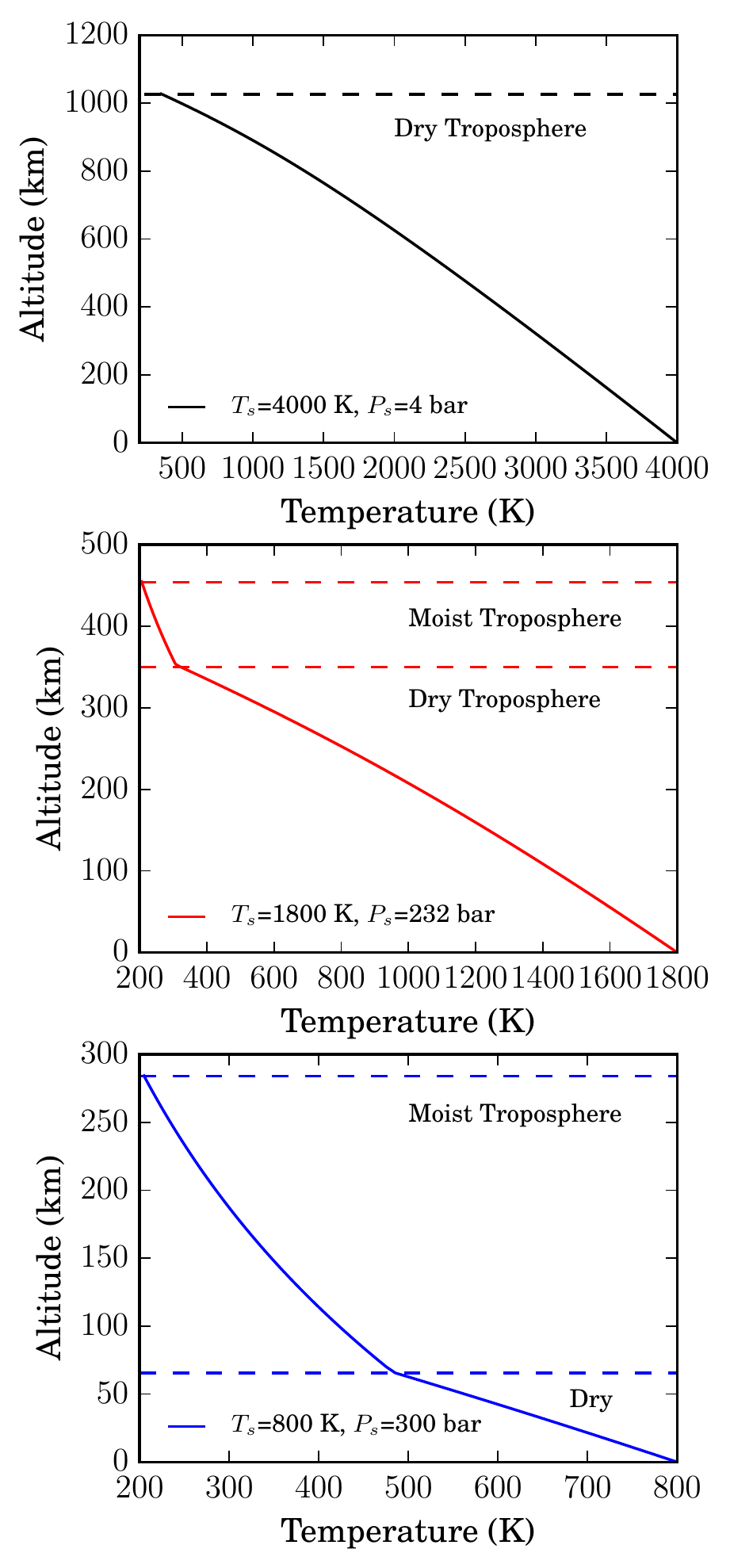}
%\plotone{figures/Ts_profiles_new}
	\caption{Temperature profiles showing the variation of altitude ($z$) (in km) vs. the temperature of the atmosphere layers for three assumed surface temperatures $T_{\text s}$ and associated surface pressure $P_{\text s}$ shown in the figure panels and are obtained as an output from the interior outgassing model assuming a grey approach. \label{f1}}
\end{figure}

%\sout{As mentioned in Sect.~\ref{int} and shown in Fig.~\ref{tp}, the outgassing rate of water is obtained as a function of the surface temperature.}
The temperature profiles for the three [$T$,$p$] pairs are shown in Fig.~\ref{f1}. Here, the upper panel illustrates that for a high temperature ($T_{\text s}= 4000$ K), a dry troposphere is present with no moist layer. However, as the surface temperature decreases ($T_{\text s}=1800$ K and $T_{\text s}=800$ K; middle panel and lower panel), moist layers of thickness 100 km and 200 km are seen, respectively. The moist upper layers are cooler than the lower layers, hence they act as a cold trap and any exchange of heat from the top of the atmosphere to space becomes less efficient.

\subsection{Atmospheric Radiative Transfer \label{lbl}}
 
For a gaseous, non-scattering atmospheric assuming local thermodynamic equilibrium, the intensity (radiance) at the Top of the Atmosphere (TOA) is given by the integral of the Schwarzschild equation
\begin{equation} \label{schwarzschild}
I( \nu ) ~=~ I_\text{b}(\nu) \, \mathrm{e}^{-\tau_\mathrm{b}(\nu)} \;+\; \int\limits_0^{\tau_\mathrm{b}(\nu)} B(\nu,T(\tau')) \; \mathrm{e}^{-\tau'} \: \mathrm{d} \tau' ~,
\end{equation}
where $I_\text{b}$ is the background radiation and $B$ is Planck's function for a black-body with temperature $T$. 
 
The monochromatic transmission $\mathcal{T}$ and optical depth $\tau$ (at position $s=0$, equivalent to $\tau=0$) is a function of wavenumber $\nu$ and path length $s$ and is described by Beer's law as follows
\begin{align} \label{beer}
\mathcal{T}\left(\nu ,s\right) ~&=~ \mathrm{e}^{-\tau(\nu ,s)} .
%  ~&=~ \exp{\left( -\int\limits_0^s \D s' \sum_m \, k_m\left( \nu,p(s'),T(s') \right) \: n_m(s') \right)} ~, \notag
\end{align}
%In the above equation, $\tau$ is written as a function of the absorption coefficient $\alpha$ 

In order to solve for the transmission of radiation (Eq.~\ref{beer}), it is important to obtain the optical depth $\tau$ which is written as:
\begin{equation} \label{absCoef}
\tau(\nu,s) ~ =~ \int_0^s \alpha(\nu,s') \, \mathrm{d}s',
\end{equation}
where, $\alpha$ is the absorption coefficient. It is usually written as the sum of the absorption cross sections $\kappa_{m}$ of the molecule, scaled by its number density $n_m$ as follows
\begin{equation} \label{alpha}
 \alpha(\nu,s) ~=~ \sum_m \kappa_m \bigl(\nu,p(s),T(s) \bigr) \, n_m(s).
\end{equation}
For high resolution lbl models, the absorption cross section of molecule $m$ is given by the superposition of many absorption lines $l$, each described by the product of a temperature-dependent line strength $S_l$ and a normalized shape function $g$ describing the broadening mechanism. The calculation for the absorption coefficient and the absorption cross section for the molecules is described in the next section.

Assuming the net energy balance of the incoming and outgoing radiation at TOA, along with the cooling flux from the magma ocean, one can write
\begin{equation}
F_{\rm OLR}- F_{\rm SUN} = F_{\rm MAG}. \label{e3}
\end{equation}

Here, $F_{\rm OLR}$ is the flux of the outgoing longwave radiation calculated using GARLIC at a viewing angle of 38$^\circ$ from zenith (See Sect.~\ref{thermal} and Appendix A) and  $F_{\rm SUN} = (1-\alpha)S_{0}/4$, where $S_{0}$ is the solar constant with 100\% luminosity  and a fixed surface albedo of $a=0.3$ is used. For simplicity, we do not include the effect of clouds and Rayleigh scattering.

\subsection{GARLIC \label{gar}}
The ``Generic Atmospheric Radiation Line-By-line (lbl) Infrared Code" or \software{GARLIC (Schreier et al. 2014)} finds its applications in several fields such as exoplanet studies \citep*{rauer11,hedelt13,Vasq13m,Vasq13c} and for Venus type atmospheres \citep*{hedelt11,Vasq13v}. GARLIC has been recently used to study Earth's atmosphere \citep*{Hoch18,Xu18} for retrieval purposes in the field of remote sensing. Moreover GARLIC has been validated using the transmission spectroscopic studies involving ACE-FTS infrared observations of the Earth \citep{Sch18a}. Furthermore, an inter-comparison of radiance obtained from the three lbl models: GARLIC, KOPRA and ARTS in 18 channels throughout the mid-IR is presented in \citet{Sch18b}.
%, we usewith temperature profiles (as shown in Fig.~\ref{f1}) as input. The model is described briefly below.

In this work, the line absorption coefficients for molecule H$_{2}$O  and other species are calculated by GARLIC using the 2010 HITEMP line list, i.e., the high temperature molecular spectroscopic database by \citet{rothman10}  and a ``CKD continuum" model \citep*{Clough89} for H$_2$O.  The latest spectral database HITRAN 2016 \citep{gordon16} is also used towards the end for comparison.  A Voigt profile \citep{Sch14}, representing the combined effect of pressure broadening together with Doppler broadening is used. 
%For the lbl computation of cross-sections, GARLIC uses a cut-off $\delta \nu=10 \rm cm^{-1}$ for H$_{2}$O with continuum.

The high resolution spectra in GARLIC are obtained by construction, i.e., the wavenumber grid spacing is adjusted to the line widths  $\delta\nu = \gamma/n$, where $n$ is by default 4 and $\gamma$ is the half width half maximum (HWHM) dependent on the pressure, temperature, and molecular mass. Since the Lorentz width is proportional to the pressure, lines become narrower in the upper atmosphere. In GARLIC, the final grid (i.e. for all layers and molecules combined) is defined by the smallest $\delta\nu$. 

We have verified the GARLIC code for a pure steam atmosphere at small temperature of  $T = 300$ K with a background surface pressure of 1 bar with the published results of \citet{Gold13} and \citet{Schaefer16}. 
\begin{figure*}[!hbt]
\centering
\includegraphics[scale=0.7]{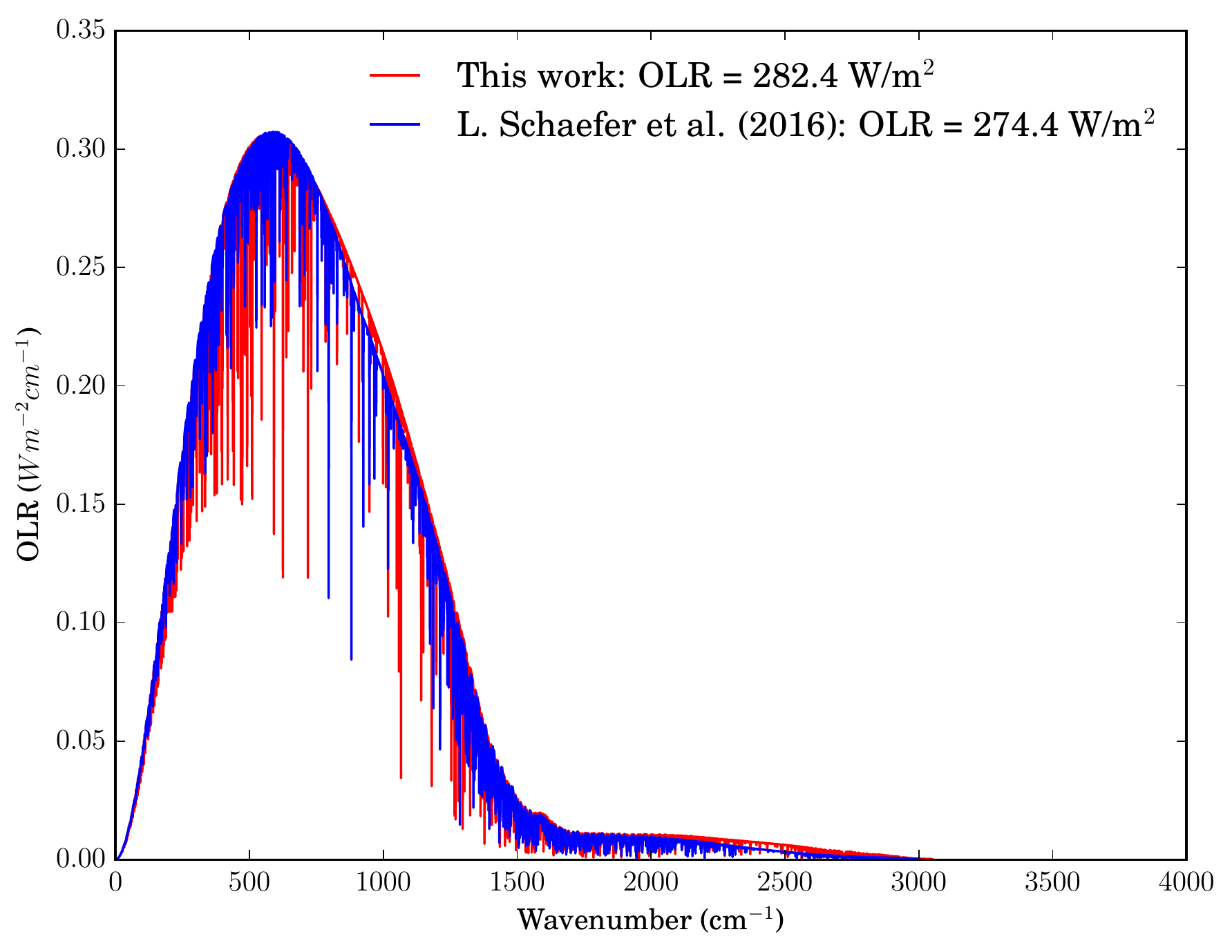}
%\plotone{OLR_NK_RW}
\caption{Comparison of emitted radiance as a function of frequency (0-3500 cm$^{-1}$) for a pure H$_{2}$O atmosphere at $T_{\text s}=300$ K and $P_{\text s} = 1$ bar using the LBL code GARLIC.   \label{f2}}
\end{figure*}
A spectrally averaged OLR of 282.4 W/m$^{2}$ is obtained which is close to the result of \citet{Schaefer16}, i.e., 274.4 W/m$^{2}$  as seen in Fig.~\ref{f2}. Also it is in a good agreement with \citet{Gold13} who obtained 282 W/m$^{2}$ for modern Earth using the LBL code SMART \citep{Meadows96}. 

At high surface temperatures and pressures and for a fixed tropopause temperature of 344 K, we compared the OLR values with the former studies using lbl codes. The comparison is presented in Table 2. The values are comparable for $T_{\text s}=1000$ K but for higher temperature $T_{\text s}=2000$ K, there seems to be a large deviation. The difference at high surface temperatures is attributed to the choice of the water continuum and due to the fact that the continuum is poorly understood at high temperatures.
        
\begin{table}[!hbt]
%\label{table2}
\centering
\caption{Comparison of OLR for high temperature runs calculated using GARLIC with the former lbl studies.}
\begin{threeparttable}
  \begin{tabular}{cccccc}
\hline\hline
	  $T_{0}$ & $T_{s}$ & $P_{s}$ & \multicolumn{3}{c}{OLR} \\ 
	  (K)  &  (K)    &   (bar) &   \multicolumn{3}{c}{(W/m$^{2}$)} \\ \hline
    344 & 1000 & 100 & & 795.9\tnote{b} & 801.4\tnote{c}\\ \hline
    344 & 1000 & 250 & 789\tnote{a} & &784.54\tnote{c}\\ \hline
	  344 & 2000 & 250 & 969\tnote{a}&    &1400\tnote{c}    
  \end{tabular}
\begin{tablenotes}
\item [a] \citet{Gold13}
\item [b]  \citet{Schaefer16}
\item [c] This study
\end{tablenotes}
\end{threeparttable}
\end{table}

% The outgassing rates are obtained from the companion paper assuming a planet with a low initial water content of 5.5 $\times$ 10$^{-2}$ wt \% (550 ppm) in the mantle and corresponding to a total of 405 bar.
\subsection{Effective height \label{eff}}
The effective height of the atmosphere, which is a function of the wavenumber describes the extent of the atmosphere and is directly measured from the transit observations. For calculating the effective height from the transmission spectra, we use the following equation
\begin{equation}
h(\nu)=\int_{0}^{\infty} \left(1-\mathcal T(\nu,z_{i}) \right) dz_{i},
\end{equation}
where the integral is calculated over all the limb transmission spectra with tangent altitude $z_{i}$ for every incident ray $i$ passing through the atmosphere terminating at the TOA. For more details, see \citet{Sch18a}.

\subsection{Chemical equilibrium with applications (CEA) \label{cea1}}
At $T_{\text s} > 2000$ K, water is expected to undergo thermal dissociation leading to a somewhat different composition of the atmosphere that may influence the OLR and thereby the MO lifetime. Therefore, we employ the NASA CEA (Chemical Equilibrium with Applications) model\footnote{\url{https://www.grc.nasa.gov/WWW/CEAWeb/}} to calculate the chemical equilibrium compositions and properties of various species formed as a result of the thermal dissociation of water at such high temperatures. The CEA calculates the chemical equilibrium concentrations of species from any set of initial composition(s) for a given [$T$,$p$]. Some of the built-in applications of the program include calculation of theoretical rocket performance, Chapman-Jouguet detonation parameters, shock tube parameters, and combustion properties \citep{Bride02,Christ16}. The CEA program utilizes a Gibbs free energy minimization routine to obtain the equilibrium-state composition for a given [$T$,$p$]. The thermodynamic data for the species are obtained from the NASA CEA database \citep{Christ16}.

The scale height of the atmosphere also depends on the composition as the atmosphere and is calculated using the same procedure as described in Sect.~\ref{str} but with an updated mean molecular weight $\mu^{*}=f_{\rm H_{2}O} \, \mu_{\rm H_{2}O}+f_{\rm H}\, \mu_{\rm H}+f_{\rm H_{2}}\, \mu_{\rm H_{2}}+f_{\rm O}\, \mu_{\rm O}+f_{\rm O_2}\, \mu_{\rm O_2}+f_{\rm OH}\, \mu_{\rm OH}$, where the additional species are formed as a result of thermal dissociation of water.

\begin{figure*}[!hbt]
\centering
\includegraphics[scale=0.6]{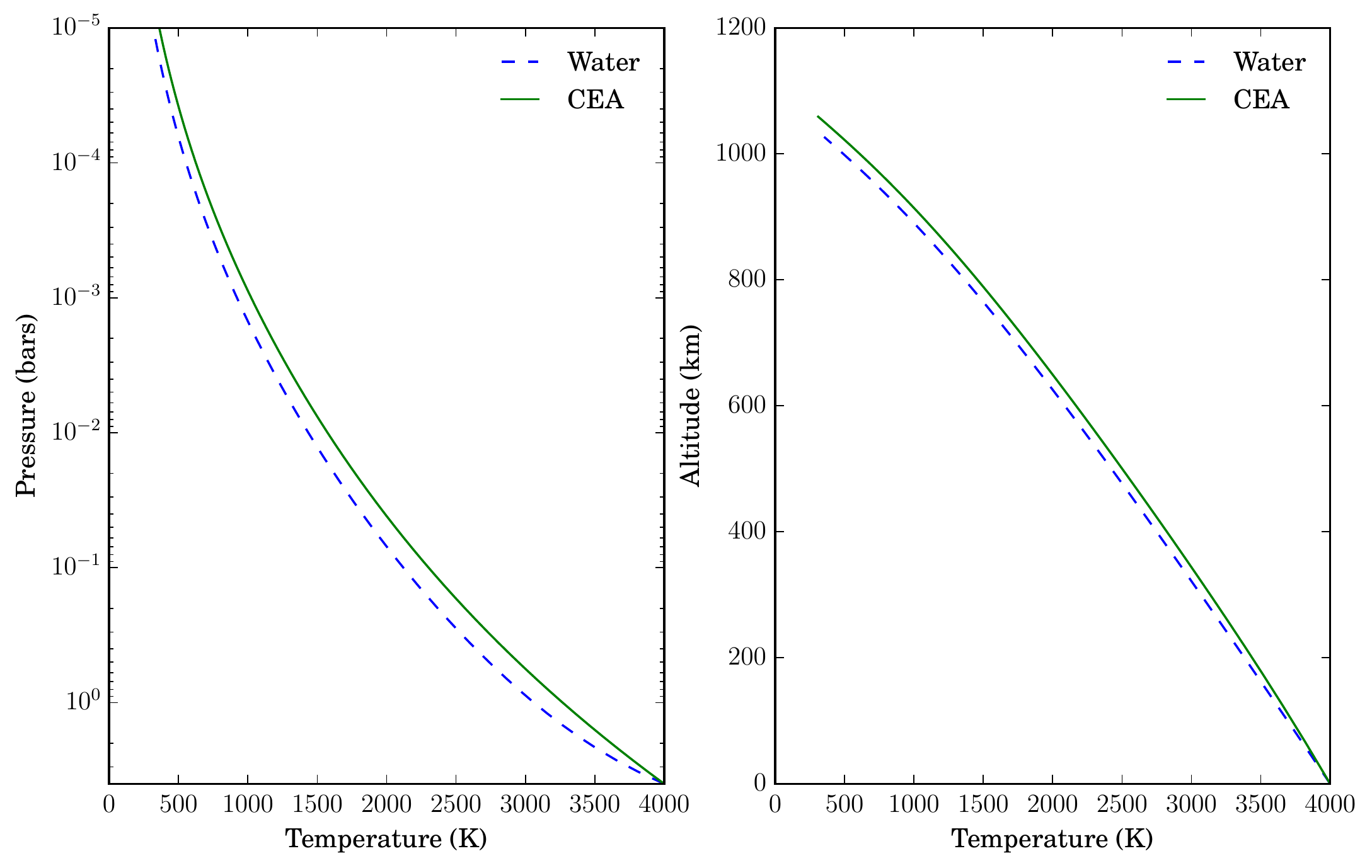}
%\plotone{figures/Ts_profiles_new}
\caption{Temperature profiles (a) showing the variation of pressure with temperature of the atmosphere layers consisting of pure water atmosphere vs. CEA for a  fixed surface temperature of 4000 K and a pressure of 4 bar  (b) altitude (z) (in km) with the temperature for pure water and CEA. \label{f1a}}
\end{figure*}

Fig.~\ref{f1a} shows a comparison between the temperature profiles of water and the CEA composition for a surface temperature of 4000 K and a surface pressure of 4 bar. Only a minor difference in the scale height is observed while comparing CEA composition with water. Also, the effect of the CEA composition on the OLR is small and is discussed later in Section~\ref{cea}.

\subsection{Framework \label{scene}}

The detailed framework of the paper is as follows. First of all, the atmospheric temperature profiles (as illustrated in Fig.~\ref{f1}) are pre-calculated for a range of surface temperatures $T_{\text s}$ = 800 to 4000 K, sampled at a resolution of $\Delta T_{\text s} = 100$ K (from 800 K - 1400 K and 1900 K - 4000 K) and a higher resolution of $\Delta T_{\text s}= \rm\, 20 K$ (1420 - 1800 K). For the surface pressure range between 4 and 300 bar, the values 4, 25, 50, 100, 150, 200 and 300 bar are chosen. In total, we calculate temperature profiles for pairs of [$T$,$p$] lying on a grid of 400 points. Then, using the temperature profiles of these 400 [$T$,$p$] points as input, $F_{\rm OLR}$ is calculated for a 100\% H$_2$O atmosphere using the LBL code GARLIC.

Secondly, the resultant $F_{\rm OLR}$ is used complimentarily to a range of MO thermal evolution experiments performed with the interior model in \citet{Nasia18}. The time evolution of surface temperature is a coupled result reported in both the studies. However, in this study, we focus on assessing the spectral evolution of the pure steam atmosphere. 

Thirdly, we obtain the outgoing longwave radiation for a special case involving thermal dissociation of H$_2$O using the CEA model. The influence of equilibrium chemistry upon the radiative spectra is also discussed.

\section{Thermal emission \label{thermal}}

%\subsection{Thermal emission \label{thermal}}

The planetary thermal emission is obtained for 400 [$T$,$p$] points as described in the Sect.~\ref{scene}. These are calculated using the LBL code GARLIC for a down-looking observer at a viewing angle of 38$^\circ$ with the zenith angle as also chosen by \citet{segura03,rauer11} and \citet{Grenfell11}. This angle is considered to be the ``mean angle" or ``equivalent latitude" at which the solar insolation incident upon Earth is 341 W/m$^2$ which is equal to one quarter of the present solar constant $S_{0}=1361$ W/m$^{2}$. The solar radiation, $F_{\rm SUN}$ in our model is taken to be one quarter of $S_0 (1-\alpha)$ which justifies the OLR calculation at 38$^\circ$ viewing angle. 
%Note that when performing coupled interior-atmosphere calculations (Sect.~\ref{couple}), the incoming solar flux is subtracted from the OLR value.

\begin{figure*}[!hbt]
\centering
\includegraphics[scale=0.3]{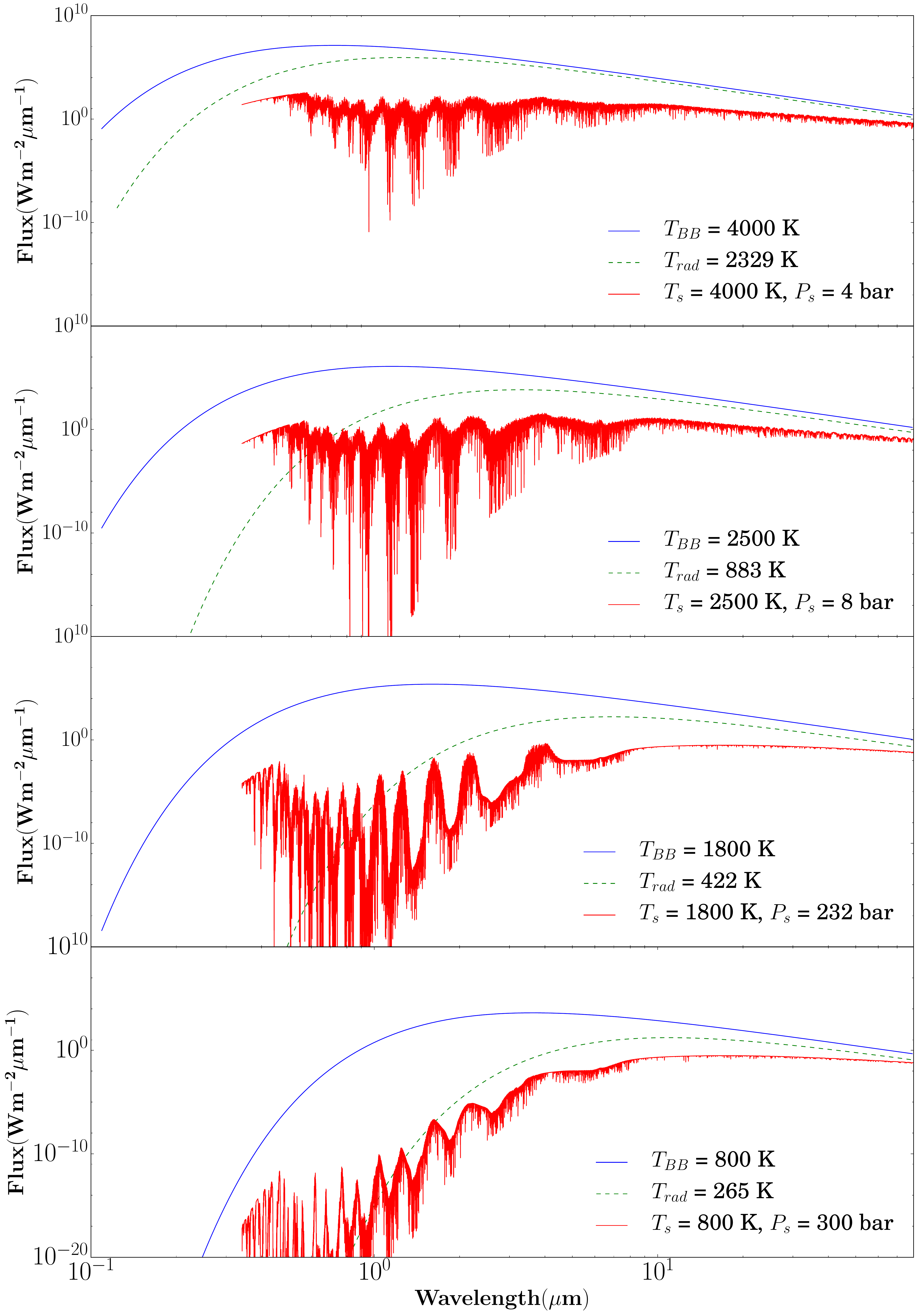}
%\plotone{figures/OLR_wave.pdf}
\caption{Emission spectra as a function of wavelength ranging from the  visible to the infrared for four different [$T$,$p$] cases. The corresponding Blackbody (BB) curves for each of these surface temperatures (in blue) and the flux at effective radiating temperature (dotted line) are shown. \label{f3}}
\end{figure*}

The flux (integral of radiance over the solid angle) in the units of $\rm\, W/m^{2} \mu m^{-1}$ at each [$T$,$p$] as a function of wavelength (ranging from mid UV to infrared) is shown in Fig.~\ref{f3} for a 100\% steam atmospheres and for four [$T$,$p$] pairs. The effective radiating temperature defined as, $T_{\rm rad}= \sqrt[4]{F_{\rm OLR}/\sigma}$ is obtained, where $F_{\rm OLR}$ is calculated by integrating the flux over 5000 points of wavenumber. Using Planck's law, the electromagnetic flux corresponding to $T_{\rm rad}$ at the radiative emitting (top) layer is shown in Fig.~\ref{f3} (dashed line). 

For a [$T$,$p$] pair with high temperature ($T_{\text s}=4000$ K) and low pressure ($P_{\text s}$ = 4 bar) as in the uppermost panel of Fig.~\ref{f3}, the peak of the maximum emission is at small wavelength (following Wien's displacement law). Hence the radiation can escape through the window regions at small wavelengths. At lower surface temperatures (lower panels), the amount of water vapour in the atmosphere increases (because of increased outgassing), leading to a decrease in the emitted flux. Also, the peak of the emission moves towards longer wavelength (following Wien's displacement law), where radiation is effectively absorbed by the atmosphere. Due to this, the spectroscopic window region becomes optically thick, e.g., at 1$\rm \mu m$. As a consequence, the radiation drops by $\sim$ 10 orders of magnitude from the uppermost panel (4000 K, 4 bar) to the lowermost panel (800 K, 300 bar).

\begin{figure}[!htb]
    \centering
        \includegraphics[width=1.0\linewidth]{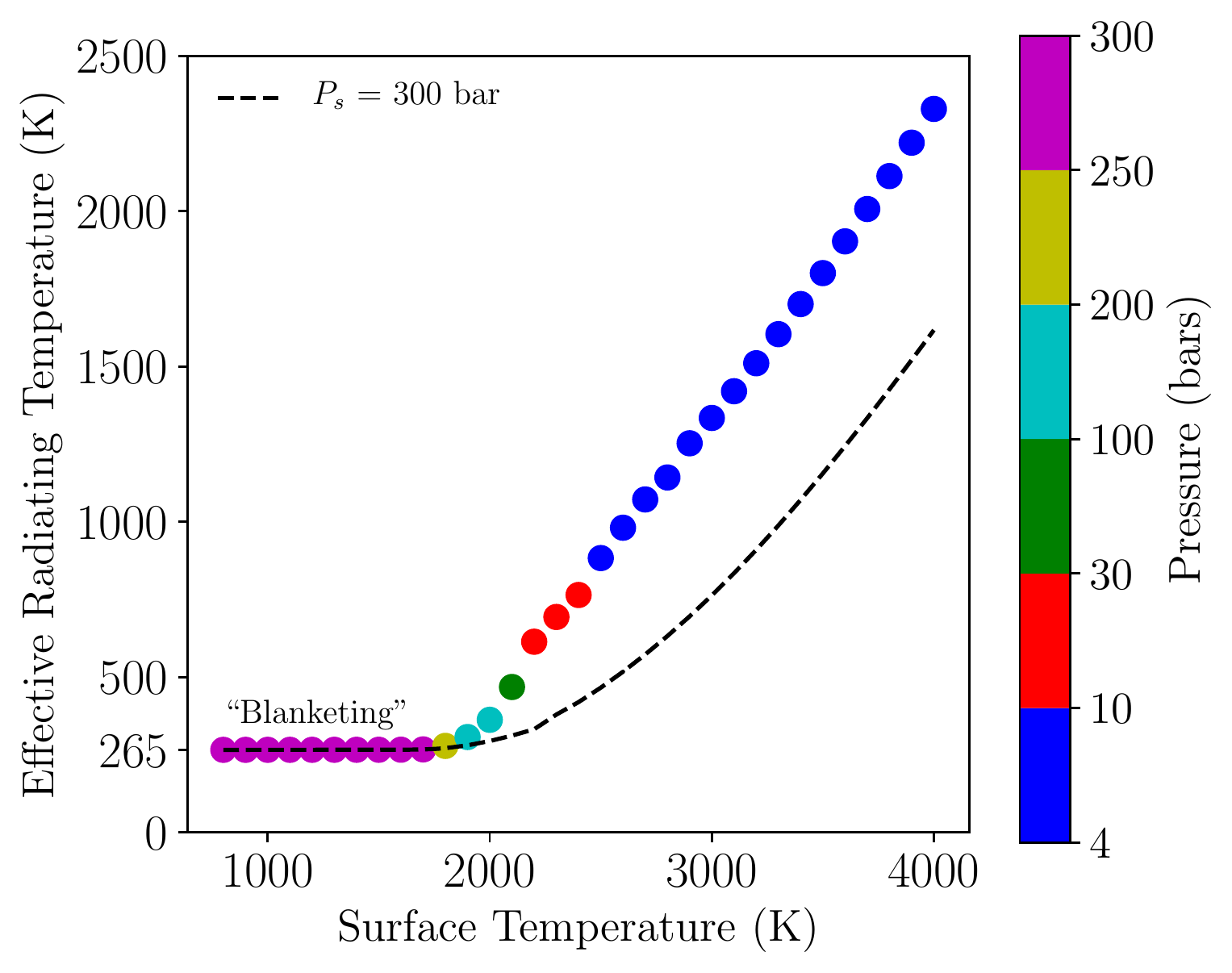}
        \caption{Variation of effective radiating temperature vs. the surface temperature with variable outgassing pressure and a fixed surface pressure. The effective radiating temperature is found to be lower for the latter case. The pressure variation is shown as the color bar. The runaway greenhouse limit is obtained for surface temperatures below 1800 K, as the surface is blanketed by water vapour. A fixed effective radiative temperature of 265 K is obtained.}
        \label{f31}
    \end{figure}

A comparison between the effective radiative temperature and the surface temperature of early Earth for variable outgassing and a fixed surface pressure of 300 bar [e.g. \citet*{Kasting88,Marcq12,Marcq17}] is shown in Fig.~\ref{f31}.

We notice here that the effective radiating temperature $T_{\rm rad}$ is less than the surface temperature, $T_{\text s}$. This is because of the greenhouse effect of water. Since we take into account a variable outgassing ranging between 4- 300 bar of water, the effective radiative temperature in the range $265 < T_{\rm rad} < 2300$ K are higher as compared to the fixed surface pressure (300 bar) where the range is  much smaller, i.e. $265 < T_{\rm rad} < \rm\, 1600 K$. For example, \citet{Marcq12} obtained an effective temperature of 350-400 K for $T_{\text s} = 2350$ K for a fixed water pressure of $\sim$ 300 bar, whereas we obtain an effective temperature of 600-700 K by considering variable pressure. In the latter case, an efficient cooling of the planet is expected. A ``blanketing effect", i.e., obscuring of the surface due to steam lying in the atmosphere begins at $\sim 1800$ K.  For $T_{\text s} < 1800$ K, the planet is effectively radiating at $T_{\rm rad} = 265$ K and a fixed OLR of $ 282 \rm \, W/m^{2}$. This means that the top of the atmosphere's thermal spectrum is bounded by the flux from effective radiation temperature of 265 K in this region. 

Using the OLR calculated for the 400 grid points of [$T$,$p$], a bilinearly interpolated grid is calculated \citep{Nasia18} to facilitate an effective coupling of the atmospheric model with the interior model within a continuous range of outgassed pressures [4,300] bar as a function of $T_{\text s}$ [800,4000] K, based on the actual [$T$,$p$] output of the interior model. 
\begin{figure}[!htb]
        \centering
        \includegraphics[width=1\linewidth]{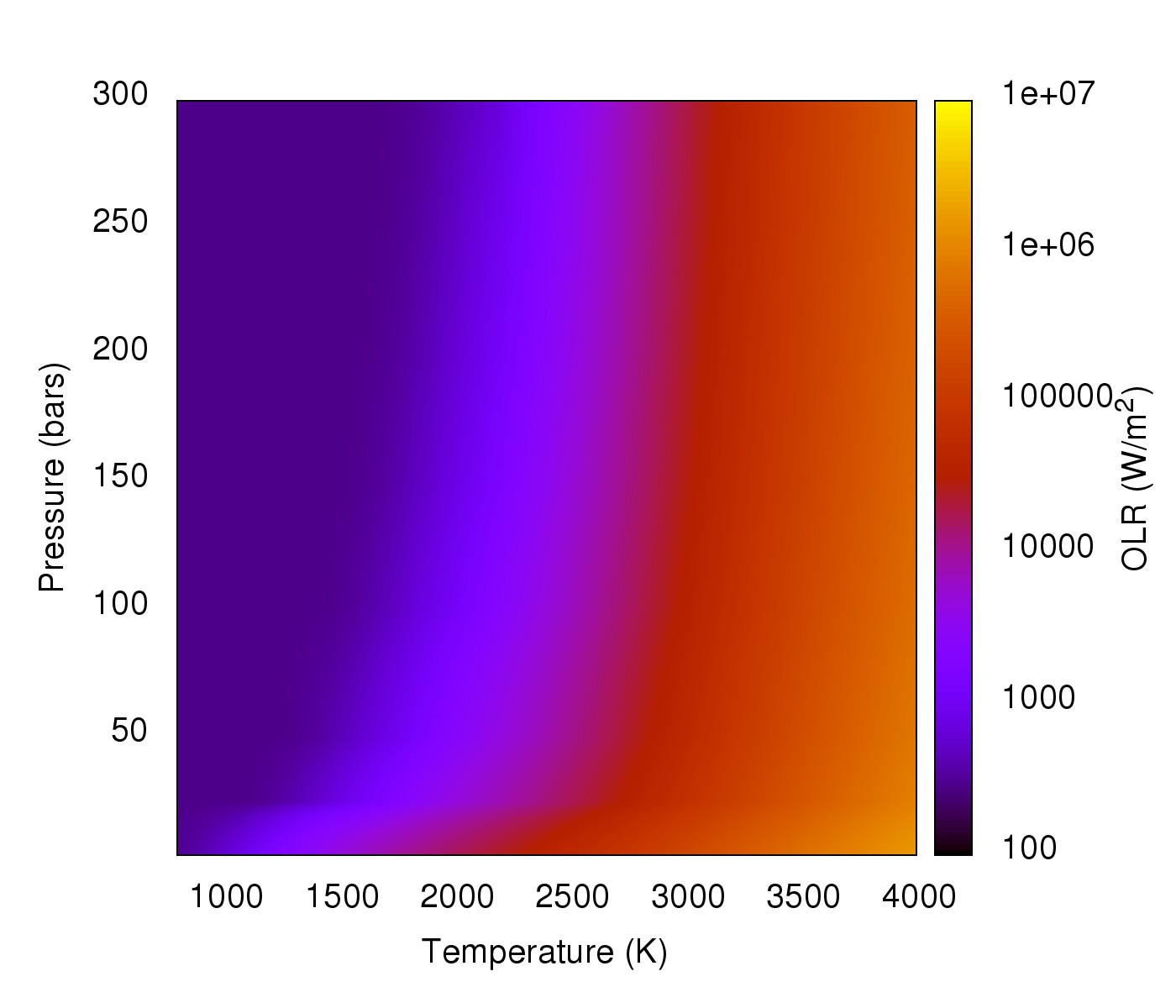}
        \caption{Result of bilinear interpolation from fixed $[T,p]$ grid of 400 points onto a new grid which then serves as an input to the interior model in order to obtain the MO thermal evolution. }
        \label{f5}
\end{figure}
The output of the interpolation is shown in Fig.~\ref{f5}. We then investigate the coupled interaction of the interior and atmosphere in the next Section~\ref{couple}. 

%\begin{figure}[!ht]
%\centering
%\includegraphics[scale=0.4]{figures/OLR_color.pdf}
%\plotone{figures/OLR_color.pdf}
%\caption{This plot shows the spectrally averaged radiance or the outgoing longwave radiation as a function for temperature for different reservoir inputs.  \label{f4}}
%\end{figure}

\section{Coupled Atmosphere-Interior model results\label{couple}}

%%As described in the previous section, the output of the atmospheric model, $F_{\rm OLR}$ is calculated on 400 surface temperature-pressure [T,p] grid points. It is then interpolated within the range 800 to 4000 K for surface temperature and 4 to 300 bar for surface pressure as displayed in Fig.~\ref{f5} using a bilinear interpolation method. 

The calculation for the MO evolutionary model (described in detail in \citet*{Nasia18}) begins at a very high surface temperature of 4000 K, which corresponds to a molten state of the full mantle. The time period of formation of Earth is assumed to be 100 Myrs. Therefore, to account for the calculation of the onset of thermal cooling of the mantle, 100 Myrs is subtracted from the time of evolution in our model. 

By assuming a radiation balance at top of the atmosphere, an iterative loop (starting at an initial time step) with a convergence criteria $\epsilon$ is run until the following fluxes are equal, meaning they numerically converge to the same value:

\begin{equation}
|[F_{\rm OLR}(T,p)- F_{\rm SUN}]- F_{\rm MAG}| < \epsilon, \label{e4}
\end{equation}        
where $\epsilon=0.1$ W/m$^2$ is equal to the uncertainty in the OLR values obtained by using GARLIC. Once the iterative loop in Eq.~(\ref{e4}) is satisfied at a particular time step, the surface temperature $T_{\text s}$ is then obtained and a new potential mantle temperature $T_{\text p}$ is calculated at the next time step by using the MO evolutionary model (described in detail in the companion paper). If Eq.~(\ref{e4}) is not satisfied, the surface temperature is adjusted by a factor $\Delta T_{\text s}$ and the corresponding $F_{\rm OLR}$ is chosen from the interpolated grid (see Fig.~\ref{f5}) and F$_{\rm MAG}$ is re-calculated. The steps are repeated until the surface temperature reaches approximately 1650 K (i.e., $T_{\rm RF,0}$), as mentioned before. Therefore, the calculations for the MO evolutionary model have been carried out until $T_{\text s} \sim 1650$ K. Note that this value is different from the $T_{\rm RF,0}$ of \citet{Hamano13}, i.e., $\sim 1370 \rm\, K$ due to different assumptions in their rheology front. We do not perform calculations for the case of a warming planet, i.e., when a negative left hand side in Eq.~(\ref{e4}) is obtained. The case for the latter is designated as planet type II by \citet{Hamano13}.

\begin{figure}[!htb]
\centering
\includegraphics[scale=0.4]{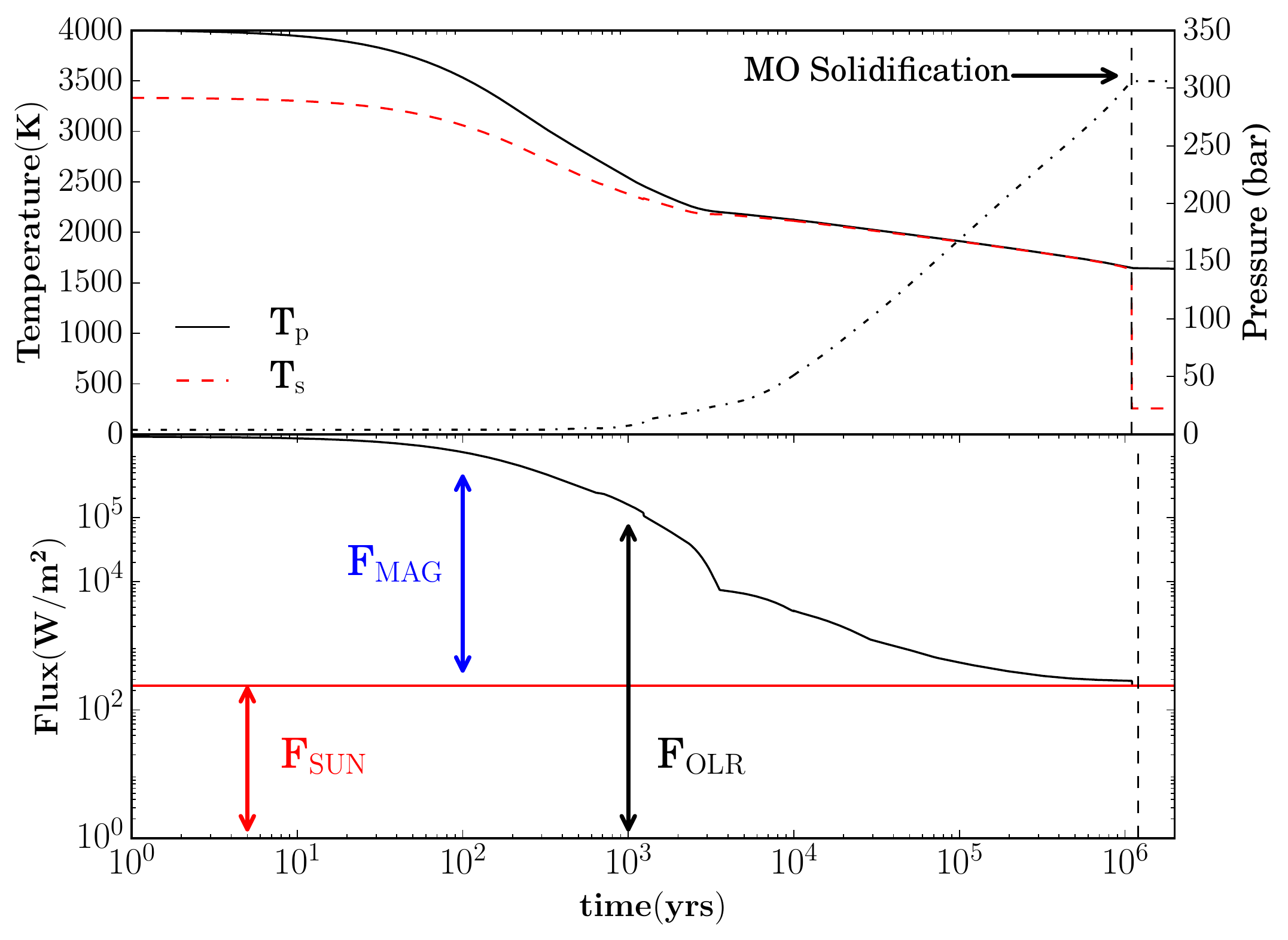}
%\plotone{figures/OLR_color.pdf}
\caption{\emph{Upper Panel}: Variation of the surface temperature $T_{\text s}$ and the potential mantle temperature $T_{\text p}$ as a function of time as obtained by \citet{Nasia18}. MO solidification (vertical dashed line) is obtained at $\sim$ 1 Myr. Evolution of surface pressure (dot-dashed line) as a measure of outgassing is also shown on the right axis. \emph{Lower panel}: Variation of the energy fluxes: flux from the emitted radiation F$_{\rm OLR}$ (black), solar incoming flux F$_{\rm SUN}$ (red) and flux from magma ocean $F_{\rm MAG}$ (blue) are shown.\label{f6}}
%, and the difference of the two i.e., $F_{\rm MAG}$ (magnitude shown by blue solid arrow). \label{f6}}
\end{figure}

The time evolution of $T_{\text s} $ is shown in Fig.~\ref{f6}. 
As seen in the upper panel of Fig.~\ref{f6}, initially both potential temperature $T_{\text p}$ and the surface temperature $T_{\text s}$ start to cool efficiently but at different rates, and reach $T_{\text s}$ $\sim 2150$ K after 0.25 $\times$ 10$^{4}$ yrs. Thereafter, cooling slows down because the atmosphere starts to limit the escape of heat flux from the interior. This is evident from the rise of outgassed steam pressure as seen towards the right of the upper panel of the figure. It can be clearly seen that the outgassed steam pressure rises from 4 bar to about 300 bar during the evolutionary phase of the MO. Finally, the MO solidification is achieved at $\sim 1 \rm\, Myr$. This is similar to the time obtained by \citet{Hamano13} when they set the surface solidification temperature of $T_{\text RF,0} \sim 1700 \rm\, K$, close to our value of 1650 K. 

Furthermore, in the lower panel of Fig.~\ref{f6}, it is seen that $F_{\rm OLR}$ approaches the radiation limit of $\sim 282 \rm\, W m^{-2}$ only by the end of the MO phase. A reduction in the $F_{\rm MAG}$ (=$F_{\rm OLR}$-$F_{\rm SUN}$) with time indicates thermal cooling of the magma ocean.

\begin{figure*}[!htb]
\centering
\includegraphics[scale=0.7]{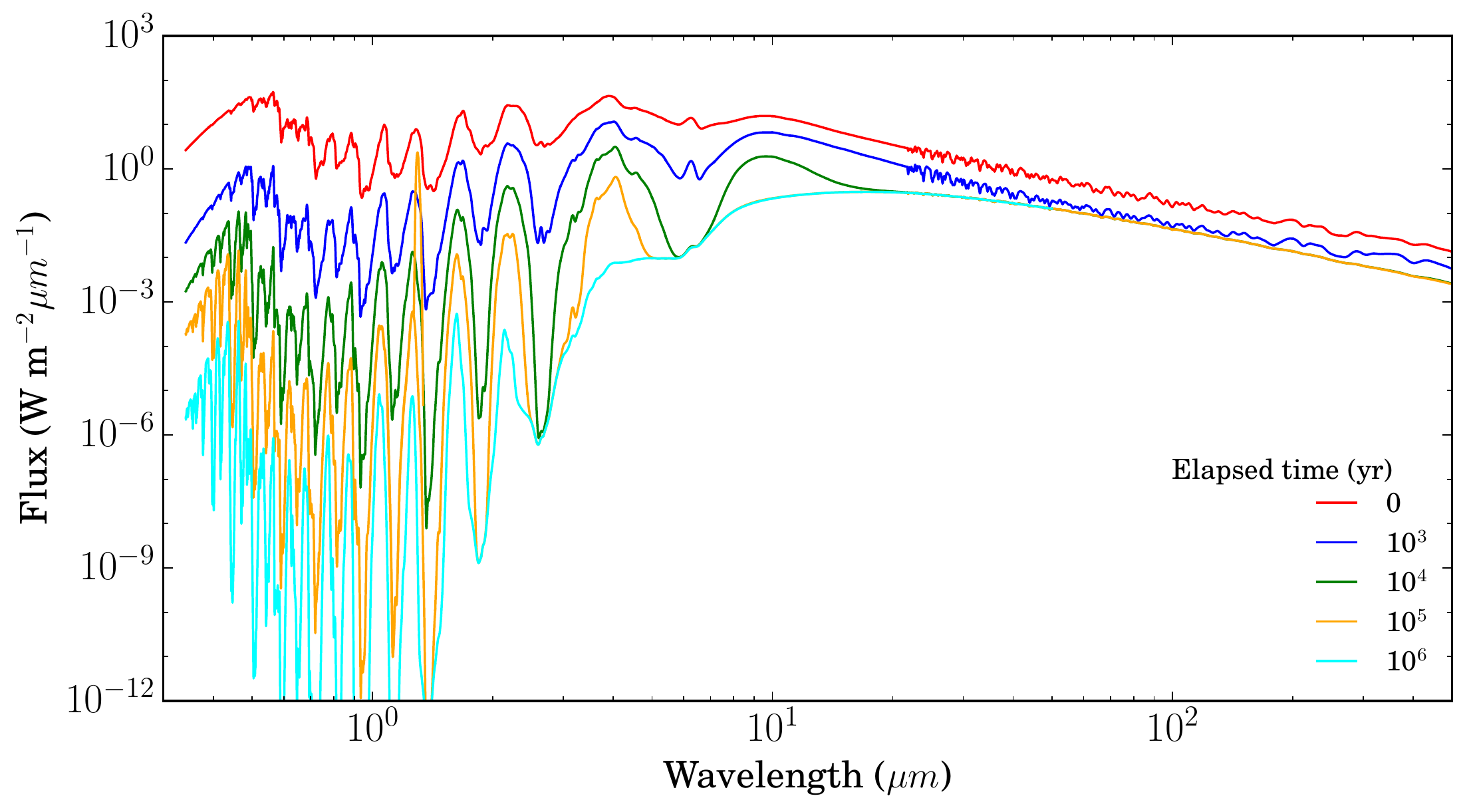}
\caption{Evolution of the thermal spectra in the visible and IR during the solidification of the magma ocean which outgasses $\sim$ 300 bars of water into the atmosphere. The spectra are shown for the entire wavelength regime 0.34-200 $\rm \mu m$ considered for the study. The color of the emission spectra indicates the elapsed time after the MO solidification starts. The thermal emission is seen to be constant for later times at IR wavelengths indicating that the emission has reached a radiation limit. \label{f7}}
\end{figure*}

Fig.~\ref{f7} shows the spectral response of the thermal emission for different times elapsed during MO evolution. For smaller timescales in early evolutionary stages (i.e., red solid line), due to high surface temperature and less surface pressure of water vapour in the atmosphere, the thermal radiation is quite high and escapes to space efficiently. However, these thermal fluxes rapidly decrease with time due to the decrease in the surface temperature and an increasing surface pressure at later times $\sim 10^{3}-10^{4}$ yr. This leads to an increased optical thickness of the atmosphere (due to water outgassing) and a `closing' of the associated IR absorption window. 

As illustrated in this figure, thermal fluxes at later times $\sim 10^{4}-10^{6} \rm\, yr$ tend to become constant for $\lambda > 10 \rm\, \mu m$  which implies that the steam atmosphere is lying in the radiation limit. This is also evident from the lower panel of Fig.~\ref{f6}, where $F_{\rm OLR}$ is seen to become constant at a value of $282 \rm\, W m^{-2}$ (OLR limit).  

The largest contribution to the emission spectra of the hottest scenarios is from the shorter wavelengths (U band) because of the absence of strong water absorption lines. In this case, a detailed diagnostic measure of a planet's atmosphere can be obtained from its thermal emission as seen in the visible and near-IR wavelengths,  i.e. M, L, Ks band (e.g.  water absorption bands in the $1-8 \rm \mu m$ region). With the decrease in $T_{\text s}$ and growth of the steam atmosphere, U band's contribution to the OLR decreases with time whereas the bands M, L, Ks initially decrease and then become constant with time indicating that the OLR limit has been reached. 

Notably, the spectral band feature at $\sim 6.3 \rm\, \mu m$ in Fig.~\ref{f7}, related to the absorption by water vapour in the atmosphere is seen to broaden (because of pressure broadening) with the increase of the amount of outgassed steam suggesting that the atmosphere is saturated with water.

\section{Observables}

\subsection{Transmission spectra \label{trans}}
Transmission spectroscopy is a widely-used tool to constrain the composition of planetary atmospheres. For exo-planetary studies, it has been recently used in this way to provide the first hints on bulk atmospheric composition and clouds of Earth-size planets, for example Trappist-1b and c \citep{Wit2016}, water bands in Neptune-sized exo-planets HAT-P11b \citep{fraine2014} and HAT-P-26b \citep{wake2017} and H$_{2}$ rich atmosphere of planet GJ3470b \citep{Ehren2014}.

\begin{figure*}[!hbt]
\centering
\includegraphics[scale=0.6]{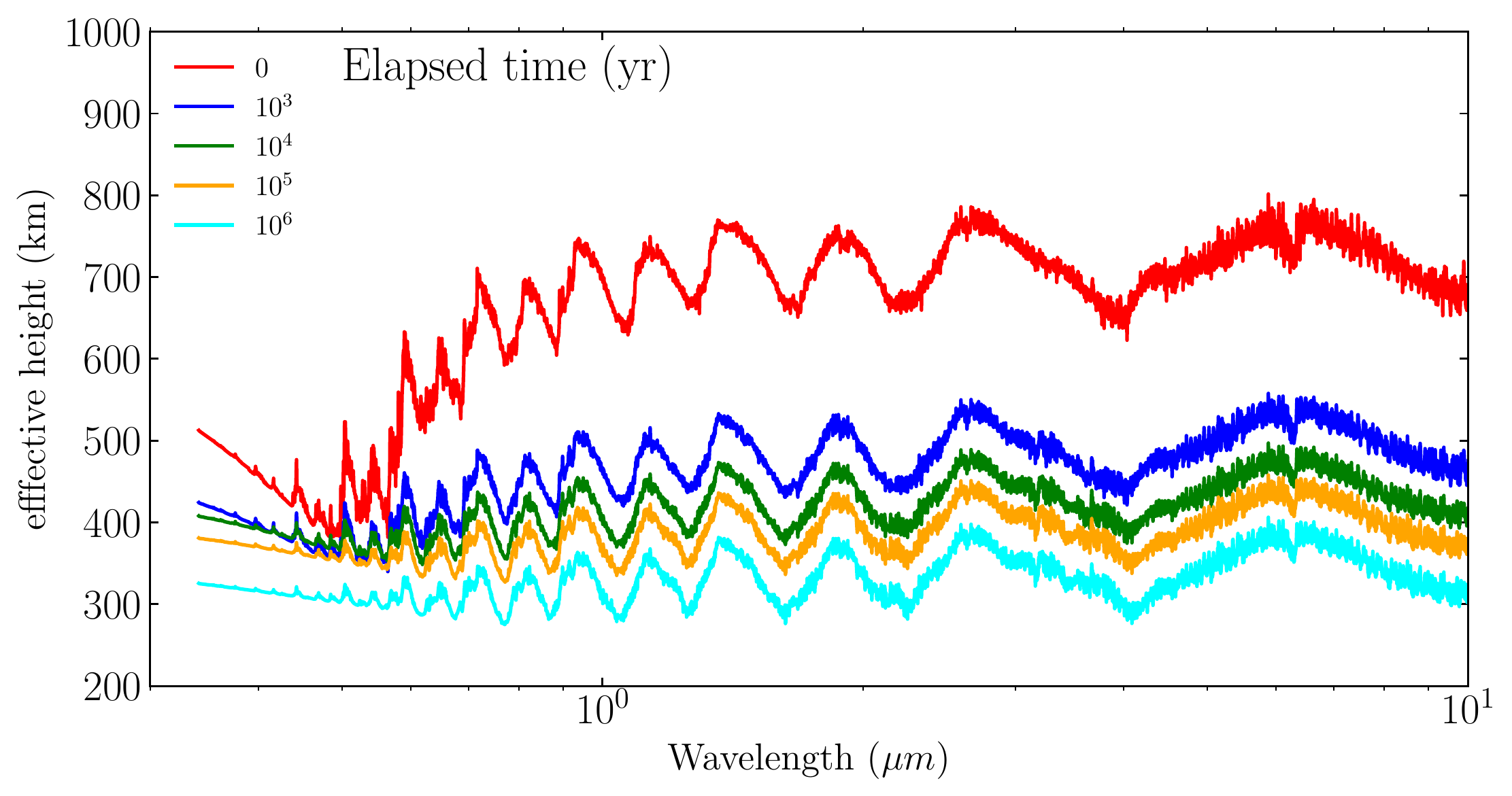}
\caption{The effective height of the atmosphere as a function of wavelength for the whole duration of magma ocean using HITEMP10 database. The effective height of the atmosphere is seen to reduce as the magma ocean solidifies.  \label{f8}}
\end{figure*}

Using our 1D atmospheres and the radiative transfer code, we calculated the transmission spectrum ranging from 0.34-30$ \rm \mu m$. It is shown as the wavelength dependent effective height in Fig.~\ref{f8} for the whole duration of the magma ocean.  The spectra are computed at a very high resolution as explained in Sect.~\ref{gar} using the HITEMP2010 database, and then binned to a resolution of $\lambda/\Delta \lambda=1000$. 
%The absorption bands of molecular water vapour are marked at the respective wavelengths, with strong features at 1.14 $\rm \mu m$, 3.3 $\rm \mu m$ and 6.2 $\rm \mu m$. 
As the magma ocean solidifies over time with the decrease in the surface temperature (see Fig.~\ref{f6}), the effective height of the atmosphere is reduced. Note that the abundance of water molecules for different elapsed time of the magma ocean is the same. The depth of the atmosphere where the absorption by the water vapor takes place is found to be $\sim$ 100 km, in contrast to 10-20 km for modern Earth \citep{Fabian18}.  

\begin{figure*}[!hbt]
\centering
\includegraphics[scale=0.6]{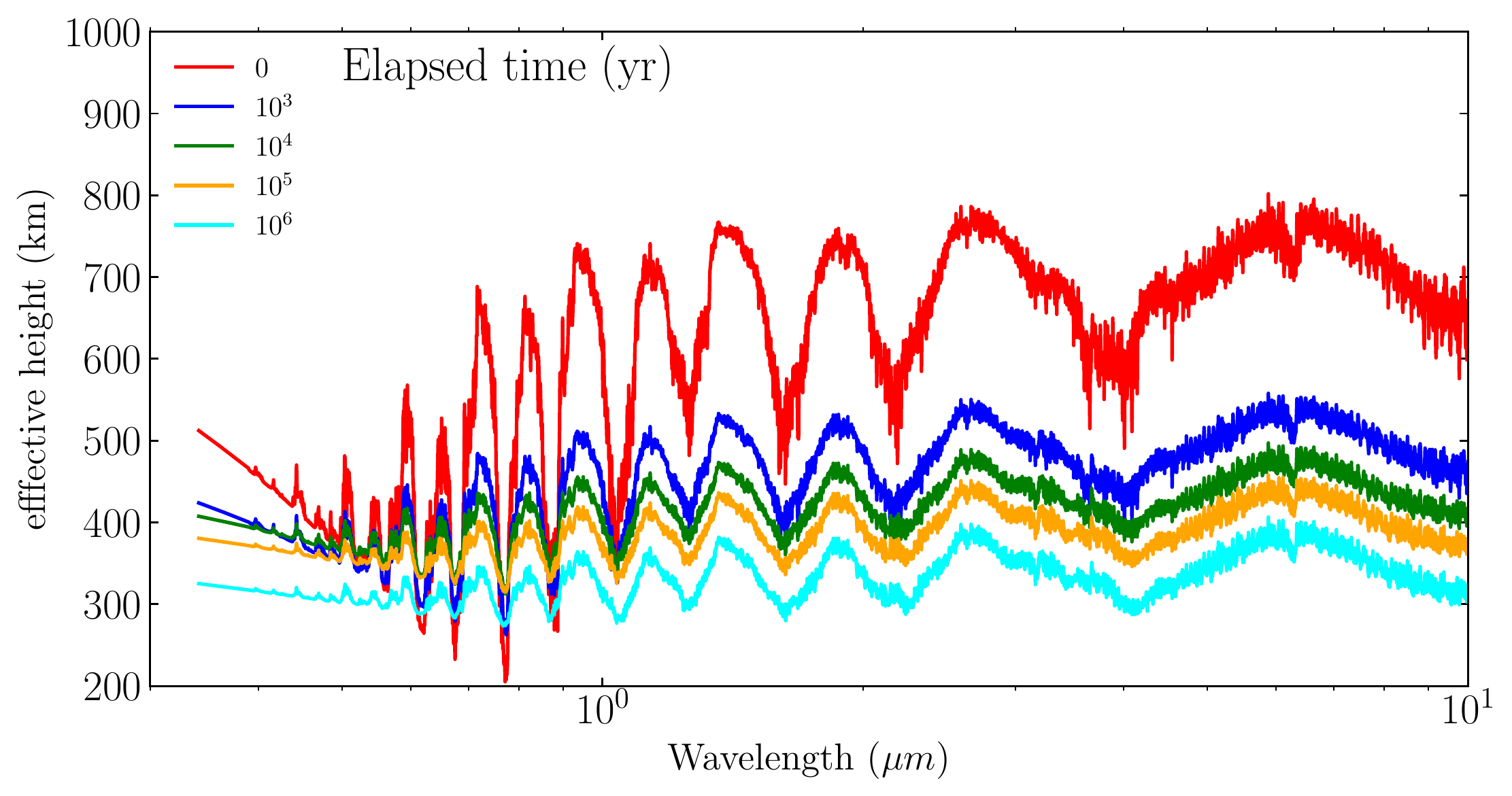}
\caption{Same as above figure but with the latest HITRAN16 database.  \label{f9}}
\end{figure*}

As a comparison, the results of the effective height using the latest database HITRAN2016 \citep{gordon16} are presented in Fig.~\ref{f9}. The variation in the absorption features in the two cases is due to the fact that HITEMP2010 consists of many more molecular bands and line transitions at high temperatures. HITRAN2016 \citep{gordon16}, on the other hand,  has an expanded database of water vapor as compared to the previous HITRAN versions. The difference in the two databases is appreciable at high temperatures and low pressures and looks same $T_{s} < 2100$ K as illustrated in the Appendix B.

\section{Chemical equilibrium Analysis \label{cea}}

\begin{figure}[!hbt]
\centering
\includegraphics[scale=0.6]{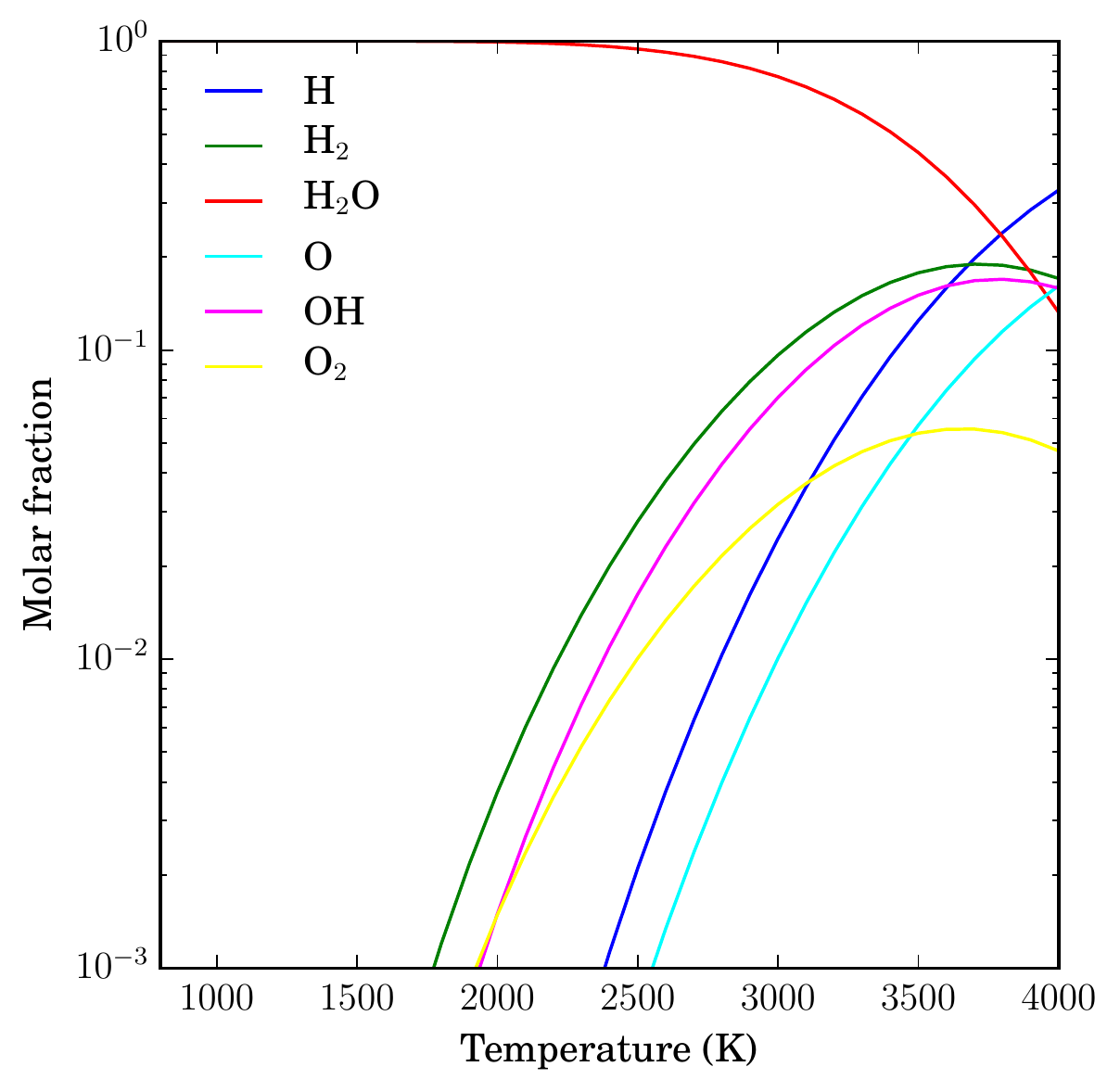}
\caption{Mole fraction as a function of temperature at pressure of 4 bar for various species formed as a result of water dissociation calculated using the NASA CEA program. \label{f11}}
\end{figure}

For the early stages of the MO, it can be expected that due to thermal dissociation of H$_{2}$O at $T_{\text s} > \rm\, 2000 K$, additional species such as H, O, O$_{2}$, O, and OH are formed. The molar fraction of these species as a function of temperature is shown in Fig.~\ref{f11}.  For this example, we set $P_{\text s} =4$ bar (corresponding to $T_{\text s}=4000 \rm\, K$ as per the interior outgassing model). It can be seen that at $T_{\text s} =4000$ K, only 13\% H$_{2}$O is present and the rest is dissociated into H (32.9\%), H$_{2}$ (17\%), O (16\%), OH (15\%), O$_{2}$ (4.7\%), and other trace elements. Moving to lower temperatures, the molar fraction of the other species reduces and water constitutes 100 \% at $T_{\text s} \sim 2000 \rm\, K$. Note there is a peak in H$_2$ and O$_2$ fractions at temperatures around $3500-4000 \rm\, K$. This is potentially relevant for assessing O$_{2}$ false positives in atmospheric biosignature science.

The molar fraction as a function of temperature obtained using the NASA CEA program are verified with another equilibrium chemistry codes, e.g. \software{IVTANTHERMO} (personal communication, L. Schaefer) which is also used in \citet{Schaefer04} and recently in \citet{SF2017}.

The higher the atmosphere, there are more chances to detect it through the pioneering transit observations. Hence, a planet in early stages of magma ocean would have perhaps more chances of detection through charaterization of its composition. 

\begin{figure*}[!hbt]
\centering
\includegraphics[scale=0.8]{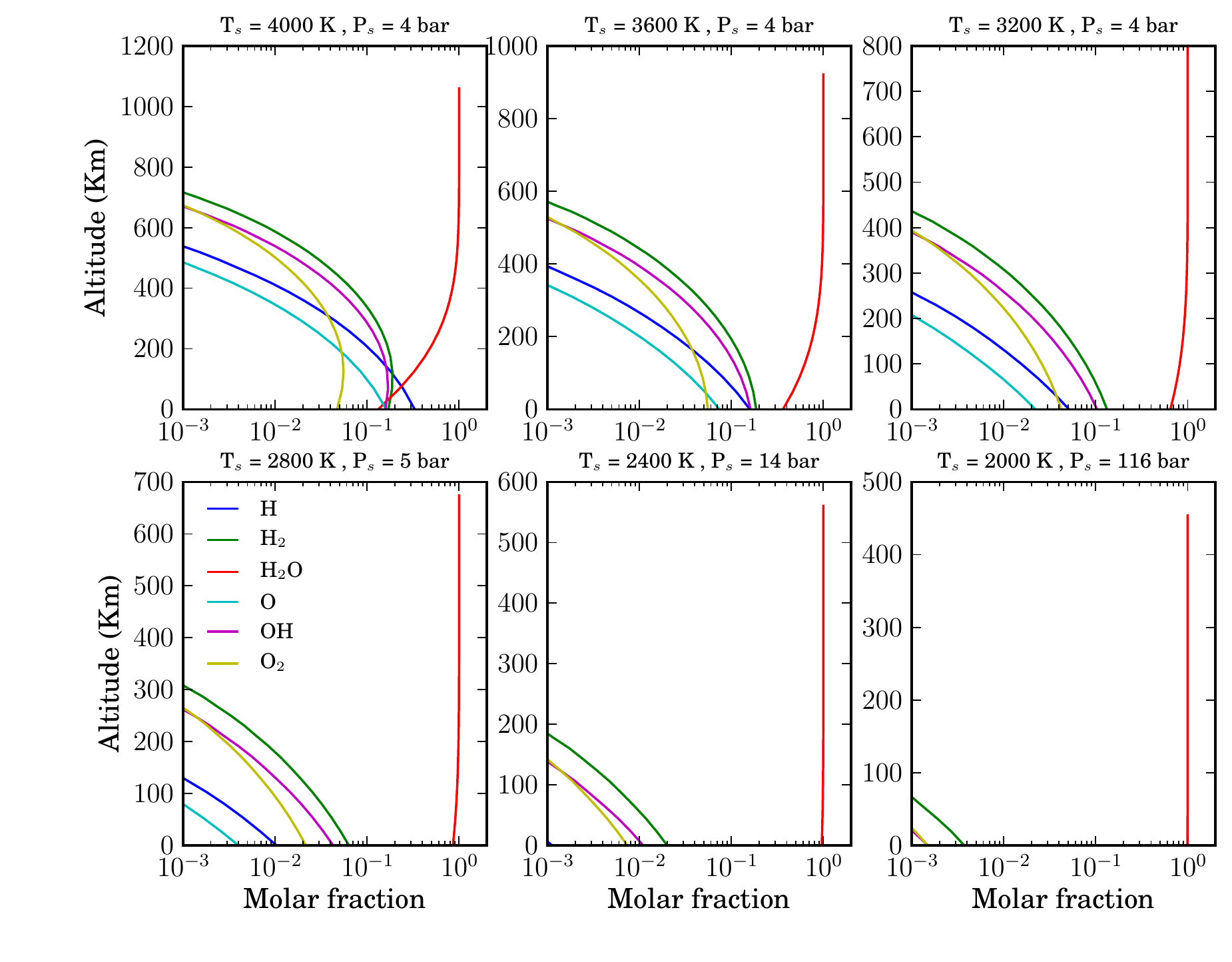}
\caption{Altitude vs. molar fraction for various species for various surface temperatures and surface pressures as indicated on top of each panel (following interior outgassing rates). This plot shows height profile of various species in the atmosphere as a result of thermal dissociation of water. \label{f12}}
\end{figure*}

Atmospheric concentration profiles (altitude vs. molar fraction) for various species  and temperature-pressure combinations are shown in Fig.~\ref{f12}. The surface temperature and pressure are according to the outgassing rates of the interior, as mentioned on the top of each panel in this figure. It is to be noted that the initial concentration of H$_{2}$O is set by the outgassing and is therefore dependent on the surface temperature.

\begin{figure*}[!hbt]
\centering
\includegraphics[scale=0.6]{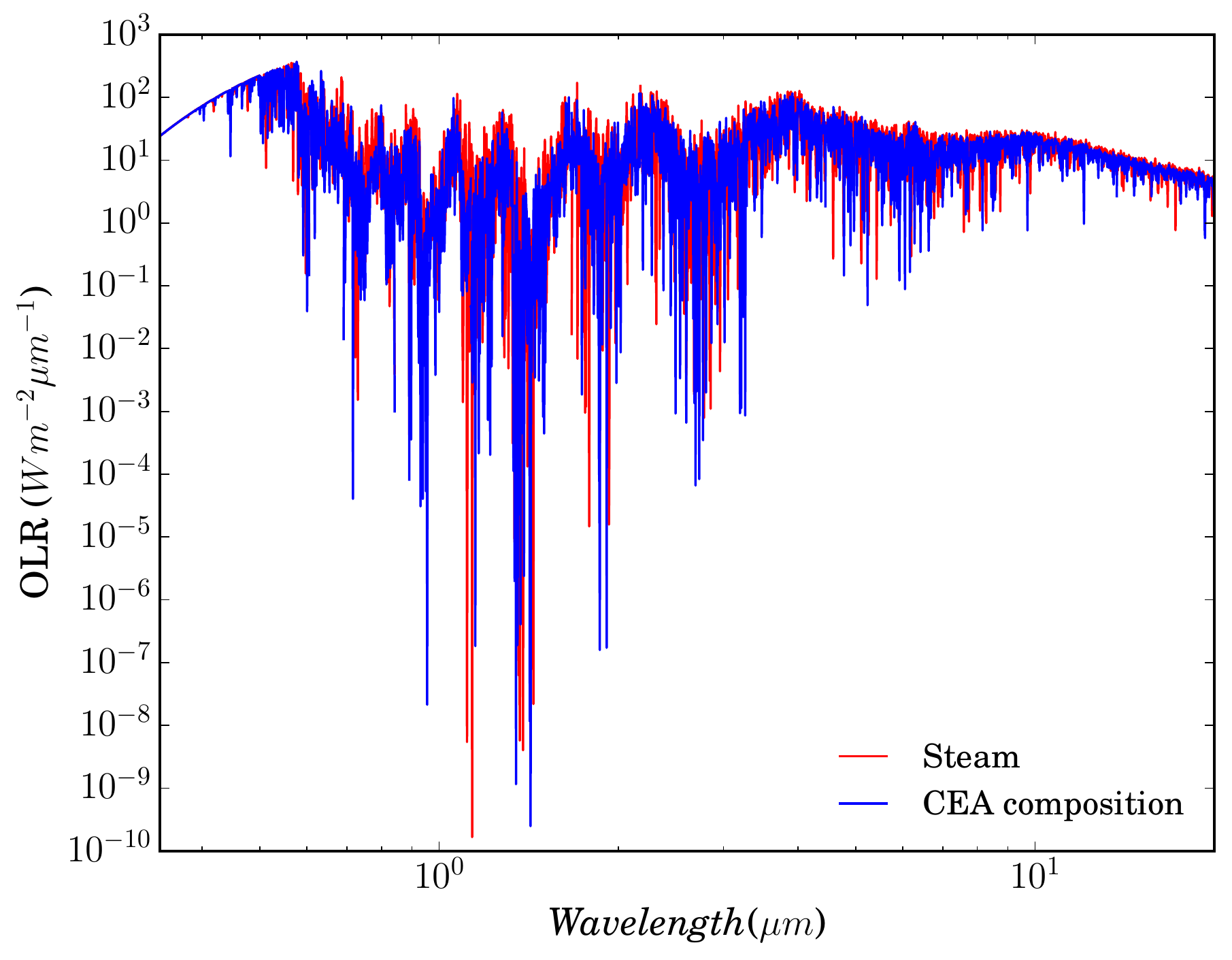}
\caption{Comparison of OLR for 100\% steam atmosphere vs. CEA composition atmosphere where the effect of thermal dissociation of H$_{2}$O is included (see text) for a case with $T_{\text s} = 4000$ K and $P_{\text s}=4$ bar. \label{f13}}
\end{figure*}

Figure~\ref{f13} shows a comparison between the emission thermal spectra or OLR as a function of wavelength for a 100 \% steam vs. the equilibrium chemistry composition (obtained using CEA) for $T_{\text s}=4000$ K and $P_{\text s}=4$ bar. The OLR for the CEA composition is found to be lower by 5-6\% than the pure steam case. The species which dominate the CEA composition namely, H$_2$ and H might be responsible for lowering the OLR. However, the O bearing species, namely O$_2$, OH and O did not result in any change in the OLR value. The impact of CEA on the MO solidifcation time is hence expected to be negligible. More discussion related to absorption by H$_{2}$ is presented in Sect.~\ref{H2}.

%%OLR calculated using the with CEA composition is shown in Fig.~\ref{f10}. OLR for CEA composition is found to be almost similar as compared to 100\% steam atmosphere as shown in Fig.~\ref{f11}. {\bf reason not CLEAR!!}  This suggests a hydrogen-dominated atmosphere with a possible greenhouse effect at the beginning of magma ocean phase (to be elaborated more based on final results). 

\begin{figure*}[!hbt]
\centering
\includegraphics[scale=0.6]{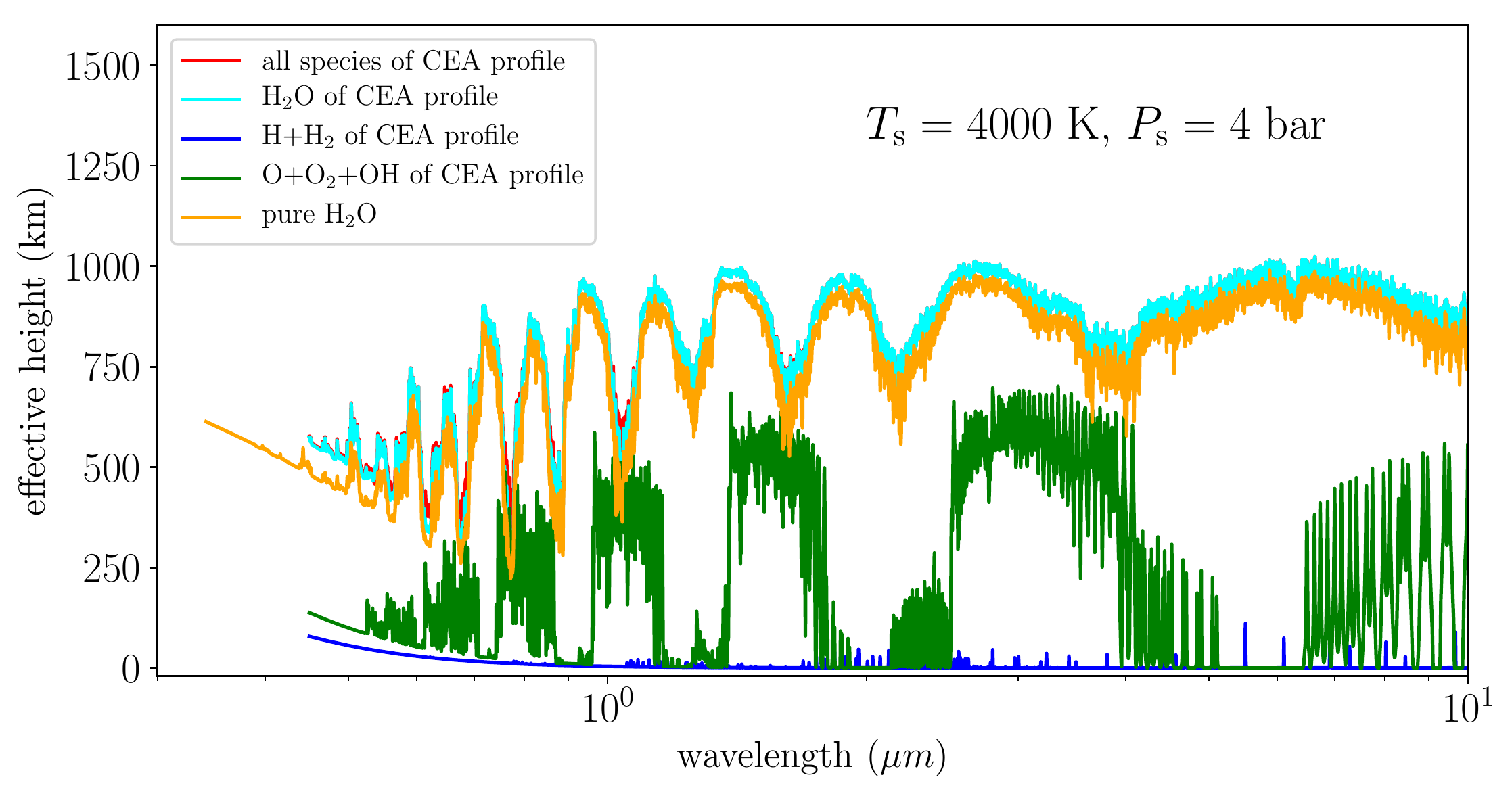}
\caption{Comparison of effective height of the atmosphere for pure water vs. individual speciation for the CEA profile using spectral database HITRAN16.  \label{f14}}
\end{figure*}

%\begin{figure}[!hbt]
%\centering
%\includegraphics[scale=0.7]{f13.pdf}
%\caption{Comparison of OLR with steam atmosphere vs. CEA atmosphere \label{f11}}
%\end{figure}
We also compare the effective height of an atmosphere constituting pure water vs. the individual species of the CEA profile at $T_{\text s}=4000 \rm\, K$ and $P_{\text s}=4$ bar using HITRAN16 spectral database as shown in Fig.~\ref{f14}. The lowered fraction of H$_{2}$O in CEA case results in a slightly higher effective height (cyan line) as compared with 100\% abundance of H$_{2}$O (orange line) according to the equation $H = k T/\mu g$. Due to the thermal dissociation of water at higher temperatures, O and H bearing species contribute a significant amount to the effective height of the atmosphere. Different absorption behavior with altitude leads to spectral features arising at different altitudes as seen in Fig.~\ref{f14}. One can see strong absorption features for the species, namely for H$_{2}$O, H$_{2}$, O$_{2}$ but the contribution of each species towards the transmission as a whole is not linearly additive. Due to this property, the effective height of the atmosphere for the CEA profile with all the species (red line) is seen to be overlapping with the CEA profile when considering only water as absorber in the lbl calculations in cyan. The difference in the spectra of H$_{2}$O of CEA profile (cyan line) and spectra of pure H$_{2}$O atmospheres (orange line) is mainly due to the increase in the scale height of the atmosphere, so the overall effective height increases.

%\sout{The presence of these additional species calculated by CEA can impact the planetary emission spectra (Fig.~\ref{f10}) and also lowers the OLR slightly for the early MO stages as shown in Fig.~\ref{f11}. As a result of the lower OLR, MO solidification time might be slightly delayed. These results are not presented in this paper and would be a part of future studies.}  

\section{Discussion \label{summ}}
One of the key objectives of the paper was to study the thermal spectral evolution of the early Earth, i.e. during the MO phase using a line-by-line atmospheric model. We have performed a systematic study of the spectral evolution for a steam based atmosphere by including the time-dependent outgassing rates obtained from \citet{Nasia18}. We briefly summarize and discuss our results below.

\subsection{Model Comparison \label{shortcomes}}
The model employed in this study has some strengths and simplifications over others. For the calculation of the thermal emission using GARLIC, we solve the radiative transfer in the spectral range 20-29,995 cm$^{-1}$ ($\sim$ 0.333- 500 $\rm \mu m$), while \citet{Gold13} have evaluated in the range of 50-100,000 cm$^{-1}$ (0.1-200 $\rm \mu m$) for a combined thermal and solar radiation calculation. For the latter, the absorption in the 0.2-0.33 $\rm \mu m$ regime is derived by using Rayleigh scattering as no LBL cross sections are available. For our purpose, we calculate only the thermal part using GARLIC with HITEMP2010 as the spectral database. The appendix provides a comparison between HITEMP2010 and HITRAN2012 \citep{rothman13}. The far-wing absorption by water vapour is a strong function of the continuum. We use a CKD approach \citep{Clough89} as compared to MTCKD2.5 used by \citet{Gold13}, \citet*{Hamano13,Hamano15}. It is to be noted that \citet{Sch18a} found the results with CKD to be almost comparable to the one obtained using MTCKD2.5. 

The specific heat capacity of water vapour has a temperature dependence. We use empirically formulated $C_{p}(T)$ provided by \citet{Wagner02} instead of using steam tables as done by \citet{Kasting88,Gold13,Marcq17} and \citet{Schaefer16}. The latter have used the steam tables by \citet{Lide2000}.

Most of the magma ocean models \citep{Elkins08,Lebrun13,Marcq17} have considered the effect of CO$_2$ along with H$_2$O as an important greenhouse atmospheric component. We have included only H$_2$O and do not include CO$_2$ in order to simplify the calculations for the atmospheric structure. Also, at high temperatures, there is a possibility of CO$_2$ getting dissociated to C and O$_2$, and both of them being sequested back to the mantle \citep{Hir2012,Schaefer16} causing the oxygen fugacity to increase. Such scenarios are hence important for outgassing by secondary volcanism during a post-magma ocean state \citep[e.g.][]{Tosi17}.  

A major strength of our atmospheric model is that it treats different amounts of H$_2$O outgassed from the interior as a function of surface temperature  as compared to \citet{Marcq17} who assumed a fixed outgassing steam pressure of 270 bar. Hence, our atmospheric model is physically more consistent with input from the interior, i.e outgassing rates as a function of varying surface temperature.

On the other hand, our model predicts a higher OLR as a function of surface temperature as compared to the OLR values of \citet{Hamano15} at fixed surface pressure. A difference of two orders of magnitude between our OLR values with the one presented in Figure 5 of \citet{Ikoma} are partly due to the non-inclusion of Rayleigh scattering process and the choice of water vapor continuum (i.e. CKD). When we included the Rayleigh scattering, this led to comparable results for fixed surface pressures as Fig. 5 of \citet{Ikoma} and \citet{Gold13} for a pure water vapour atmosphere. An illustration of differences in the OLR values using several different radiative transfer codes for steam-based atmospheres is presented in \citet{Yang16}.

Furthermore, since \citet{Marcq17} have included spectral wavelength longward of 1 $\rm \mu m$ only for the calculation, they might be underestimating the OLR values for $T_{\text s} > 2400 \rm\, K$, wherein most of the absorption happens at wavelength shortward of 1 $\rm \mu m$. Therefore, in our study, we have extended the shortwave calculations up to $0.34 \rm\, \mu m $ (29,445 cm$^{-1}$). As a result, we find that the component of OLR shortward of $1 \rm\, \mu m$ has an important contribution, i.e. it  contributes $\sim$ 1.2\% of the total OLR for $T_{\text s}=1800 \rm\, K$; 38\% for $T_{\text s}=2500$ K and $\sim 90$\% for $T_{\text s}=4000 \rm\, K$.  This suggests it is important to include a radiation contribution shortward of $1 \rm\, \mu m$ to account for the absorption of thermal radiation and outgoing longwave radiation at high surface temperatures (e.g., $T_{\text s} > 2400$ K). 
%Note that by including Rayleigh scattering, a reduction in OLR is obtained.

In this paper, we have not included atmospheric escape of hydrogen and oxygen as illustrated by e.g., \citet{Schaefer16,SF2017} but we included the thermal dissociation of water at higher temperatures leading to the formation of additional species such as H, H$_{2}$, O, O$_{2}$, and OH in chemical equilibrium with each other.  Out of these, H$_{2}$ could act as a greenhouse gas which is lying in the lower parts of the atmosphere. This results in lowering of the OLR by causing more absorption of radiation in the infrared spectral regime. In addition, H$_{2}$O may get photolysed, which may lead to a further change in composition of the atmosphere. 

%Also, the formation of O$_{2}$ at higher temperatures can be linked to the abiotic production of O$_2$. 

\citet{Hamano13} have found a significant water loss of type II planet during the magma-ocean period which lasted around about 100 million years (long-term MO). They have also suggested a strong escape of H-atoms to space while O$_2$ would be dissolved in the MO. The latter is believed to increase the oxygen fugacity of the mantle by up to about 0.07 wt\%.
 Therefore, the photolysis and H escape, both are significant processes that impact the atmospheric composition and mass, for e.g. it is also  shown by \citet{Schaefer16} that H escape leads to the build up of O$_{2}$. The planet GJ1132b studied by \citet{Schaefer16} is in a long term MO, so H escape is a significant process in order to loose more water quickly. 
In contrast, for a short duration MO planet studied by us, we only studied the process of thermal dissociation of water which leads to a H$_{2}$-dominated lower atmosphere along with some abiotic production of O$_2$ during the onset of the MO. This corroborates with the findings for the case of exoplanet GJ1132b by \citet{Schaefer16} and recently for the two innermost TRAPPIST planets- TRAPPIST-1b,1c \citep{SF2017}, that employed chemistry calculations along with diffusion.

\subsection{MO solidification times}
The initial content of H$_{2}$O inside the mantle is assumed to be low, i.e., 5.5 $\times$ 10$^{-2}$ wt \% in this study. Using this, we find that the melt fraction at the surface reaches the critical value of 0.4 at a temperature $T_{\text s} = 1650$ K, marking the end of the magma ocean where the entire mantle exhibits a solid-like behavior \citep[e.g.][]{Nasia18}. Hence, the mantle cools down via solid-state convection much more slowly than via liquid-like convection, which characterizes the magma ocean phase. We estimate the solidification time of the MO at this surface temperature to be $\sim 1 \rm\, Myr$. \citet{Hamano13} have obtained the duration of magma ocean to be $\sim 4 \rm\, Myrs$ for a surface solidus temperature of $T_{\rm s}= 1370\rm\, K$ (using a different rheology front) and for an increased initial water inventory (5 $M_{\rm EO}$). However, their estimate becomes quite close to our findings upon setting their solidification temperature to  $T_{\text s}= 1700 \rm\, K$ (close to our value of 1650 K) in which case the lifetime of magma ocean reduces to $1 \rm\, Myr$ (see also discussion in \citet*{Nasia18}).

Furthermore, upon using the original setup of the \citet{Marcq12} atmospheric model, the duration of MO solidification is reported to be $\sim$ $1 \rm\, Myr$ in \citet{Lebrun13}. However, the updated model version of \citet{Marcq17} obtained a 2.5 times reduction in this value, i.e.to $8.0 \times 10^{5} \rm\, yrs$ for a H$_2$O-CO$_2$ dominated atmosphere, and for a slightly smaller initial water inventory, i.e. $4.3 \times 10^{-2} \rm\, wt \%$. On the other hand, a higher initial H$_2$O inventory of 100 Earth oceans used by \citet{Schaefer16} is able to extend the MO lifetime to $\sim 8 \rm\, Myr$ due to enhanced outgassing. Hence, the solidification timescales are very sensitive to the initial water inventory used. These along with the other factors affecting the MO solidification times are discussed in detail by \citet{Nasia18}.

Our results of MO solidification timescale compare well with the existing 1D radiative-convective model studies presented by \citet{Hamano13,Hamano15} and \citet{Marcq17} for similar parameters. In Table 3, we provide a list of various coupled atmospheric-interior models in the literature and present a comparison of their results with our study.

\begin{table*}[!hbt]
\small
\label{table2}
\centering
\caption{A comparison table for the MO cooling timescales for various coupled atmospheric-interior models. The MO cooling timescales are mentioned in the last column with the type of planet I or II and the water inventory in the parentheses.  }
\begin{tabular}{|l |c |c|c|c|}
\hline\hline
Reference & Rad. transfer  & Initial water inventory & Solidification  &  MO cooling timescale  \\
          &  method              & M$_{\rm EO}$ \& (wt \%)    & temperature  (K)                &         (Myr)     \\ 
\hline
\citet{Lebrun13}  & correlated-$k$ & 1     & 1,400     &      1 (Planet Earth)             \\
                  &   (grey)             &   (4.3$\times$ 10$^{-2}$)   &   &\\
\citet{Hamano13}  & correlated-$k$ & 0.01-10   & 1,370    &  4 (Type I; 5 M$_{\rm EO}$)            \\
                  &   (grey)             & 31,370    &  100 (Type II)            \\
\citet{Hamano15}  & correlated-$k$ & 0.01-50    &1,370   &3.9  (Type I; 5 M$_{\rm EO}$)           \\
                  &   (non-grey)             & (2.3$\times$ 10$^{-4}$-1.2)     &1,370   &100 (Type II; 5 M$_{\rm EO}$)\\
                  &                &      &1,700   &1 (Type I)\\
\citet{Schaefer16}& LBL            &0.1-1000 & 1,420    &       8 (100 M$_{\rm EO}$)            \\
                  &   (non-grey)             &  (up to 20)    &   &\\
\citet{Marcq17}   & correlated-$k$ & 1     & 1400-1,650    &  0.8            \\
                  &                &(4.3 $\times$ 10$^{-2}$)      &   &\\
This study (2018)       & LBL GARLIC     &  1   & 1,650    &  1        \\
                  &   (non-grey)             & (5.5 $\times$ 10$^{-2}$)     &   &\\
 \hline
\end{tabular}
\end{table*} 

%%Discussion about critical albedo 0.14 and how permanent MO state could be achieved. 

%\begin{figure*}[!hbt]
%\centering
%\includegraphics[scale=0.6]{eff_heighti_CEA.pdf}
%\caption{Comparison of effective height of the atmosphere for pure water vs. CEA case. A higher effective height for the CEA case suggests more absorption. \label{f12n}}
%\end{figure*}

\subsection{H$_{2}$ as greenhouse gas \label{H2}}
H$_{2}$ has been discussed as an important infrared absorber and radiative gas by e.g., \citet{PG11} and \citet{RW12} who suggested warming of the Martian surface due to an enhanced H$_2$ greenhouse effect for at least a few tens of Myr. Using a 1-D climate model for early Mars with a surface temperature of 273 K and pressure of 2 bar, \citet{Ram14} have obtained a reduction in the OLR by 6 and $22 \rm \, W m^{-2}$ due to the increase in the H$_2$ abundance by 5 and 20\%, respectively.

Diatomic H$_{2}$ interacts with the radiation via collision-induced absorption (CIA), absorbing strongly in the middle and lower portions of the atmosphere and causing it to be a potential contributor to greenhouse warming during the early Earth \citep{robin2013}. Moreover, CIA by H$_2$-H$_2$ is known to dominate the radiative transfer of heat in the atmosphere of early Earth and other major planets. For example, a study by \citet{mckay} shows H$_2$ as one of the key absorber of radiation in the Titan's infrared spectrum and hence responsible for its greenhouse effect. 

Our results on the thermal dissociation of steam using chemical equilibrium runs suggest a small build-up of O bearing species: O$_2$, OH, and O. On the other hand, a significant fraction of H (32.9\%) and H$_{2}$ (17\%) is found for the case with surface temperature of $T_{\text s}=4000$ K (i.e. at the onset of MO solidification). The concentration of these newly formed species, however, decreases as the temperature decreases. Looking at the atmospheric concentration profiles (Fig.~\ref{f12}), we see that the newly formed species with a molar fraction $> 10^{-2}$ are lying at altitude less than 600 km in the atmosphere, and a lesser fraction of molecules ($< 10^{-3}$) is lying above 600 km. Hence, the radiation is absorbed by H$_2$ in mostly the lower and middle layers of the atmosphere, i.e. the infrared part of the spectra. To corroborate the above fact, the results on the contribution of effective height of the atmosphere due to individual species show H$_{2}$ as an absorber in the lower part of the atmosphere. As a result, a higher H$_2$ and H concentration in the lower and intermediate atmospheric layers could possibly enhance the warming due to greenhouse effect. Moreover, we obtain a lower OLR by 5.8\% (for  $T_{\text s}=4000$ K) and 1-5\% (for $3500  > T_{\text  s} > 4000 \rm\, K$) in the case of CEA as compared with steam, because of the absorption caused by H$_{2}$. Note that we did not estimate the changes in the OLR due to hydrogen escape. Clearly, a consistent model of atmospheric escape is required to investigate this issue further.

\section{Summary and Conclusions \label{Conc1}}
In this paper, we have investigated the evolution of the steam atmosphere during the magma ocean period of the early Earth using the atmospheric lbl code GARLIC based on a non-grey approach. The evolution of the atmosphere is invariably linked to the thermal cooling of the magma ocean. We assumed a time-dependent outgassing of water vapour, leading to a variable outgassed pressure ranging between 4 and 300 bar and surface temperatures between 800-4000 K. Using variable outgassed pressure instead of choosing a constant value, i.e. 270 bar, we  determine higher estimates for the effective radiating temperature, as compared to the previous studies. In this paper, we have studied the atmospheric spectral evolution  by calculating the outgoing longwave radiation and the effective height of the atmosphere as a function of MO cooling timescales. The thermal dissociation of water at high temperatures and its effect on the OLR and effective height of the atmosphere is also presented. The main conclusions of our study are listed below. 

(1) For relatively hot surfaces ($T_{\text s} > 2500 \rm\, K$), a larger flux of radiation leaves the planet due to a smaller pressure of water on the surface. This leads to a significant and rapid cooling of the magma ocean. We find out that such atmospheres usually have high effective radiating temperatures ($ \rm\, 1000-2300 \rm\, K$) and hence are easier to characterize observationally. 

(2) For relatively cool surface temperatures ($T_{\text s} < 2500 \rm\, K$), a limited flux is able to escape to space, with moderate effective radiating temperatures ($\rm\, 265-800 \rm\, K$) and a strong blanketing of the surface is seen. Such atmospheres are difficult to observe probably due to the thick overlying layer of clouds. 

(3) Due to the blanketing effect of water, only a limited amount of radiation can escape to space in order to cool the planet. This limiting value of the outgoing longwave radiation, known as the OLR limit is estimated to be $282 \pm 2 \rm\, W m^{-2}$ using the lbl code GARLIC. This value is consistent with previous studies that reported a range between 280 to $310 \rm\, W m^{-2}$ \citep*{Kasting88,Abe88,Kopp13,Gold13,Hamano13,Marcq17}. At this stage, the planet is in a runaway greenhouse regime, radiating at an effective temperature of 265 K. Moreover, we conclude that the difference between the surface temperature and the effective radiating temperature for a planet is a strong function of its surface pressure and is important for estimating the radiative heating due to the greenhouse gases present in the atmosphere.

(4) By coupling the climate model (this work) and the interior model (companion paper), the solidification timescale for the MO is found out to be $1 \rm\, Myr$ at a solidification temperature of $1650 \rm\, K$, assuming a pure steam atmosphere derived from initial water reservoir equivalent to one Earth's ocean. One of the main application of our work would be to compare the results of the lbl model ``GARLIC" (non-grey) used in this study with the grey approach \citep{Lebrun13}. This has been done by \citet{Nasia18}, who have found a delay in the MO cooling timescale by few hundred thousand years by using results of our lbl (non-grey) atmospheric model as compared to a grey model.  Furthermore, they have also discussed several other factors responsible for the delaying of the MO timescale.

(5) The transmission spectra for the whole time duration of the magma ocean for pure water atmospheres is presented as wavelength-dependent effective height yielding $\sim$ 100 km depth of the atmosphere, where the absorption by water molecule takes place. The effective height of the atmosphere decreases as the magma ocean solidifies or the surface temperature reduces.

(6) Finally, we have included the thermal dissociation of H$_{2}$O at T$_{\text s} > 2000 \rm\, K$ calculated using chemical equilibrium analysis (CEA) and studied its effect on the OLR. We suggest a H$_{2}$ rich early Earth's atmosphere with about $\sim 10-30$\% volume fraction for surface temperature ranging between $3500-4000 \rm\, K$, respectively, i.e. during the initial evolutionary phase of the MO. It is responsible for lowering the OLR by 1-6\% as compared to the pure water case. The thermal dissociation also resulted in an abiotic formation of O$_{2}$. H$_2$ formed in the atmosphere may slowly
diffuse through the atmosphere and hence be transported to the
higher layers and escape. Photo-dissociation of
H$_{2}$O, which is yet another important process for H$_{2}$ formation and escape in these atmospheres, has not
been included in this study. For the CEA profile, we also studied the transmission spectra. In this spectra (Fig.~\ref{f14}), the thermal dissociation of water into H and O bearing species led to an increase in the effective height of the atmosphere and can be clearly seen in the beginning of the MO ($T_{\text s}=4000$ K and $P_{\text s}$ = 4 bar), which decreases as the surface temperature cools down.

%Due to the thermal dissociation of water, H$_{2}$ formed in the lower atmosphere might slowly diffuse to the upper layers and hence be transported to higher layers and (slowly) escape. The escape process of H and the photo-dissociation of H$_{2}$O, which are yet another important physical processes, have not been conducted in this study. 

%(6) We find that the CEA composition constituting of H$_{2}$ (as a result of thermal dissociation) during the early Earth period is responsible for absorbing $1-6$\% of radiation, causing warming and a reduced OLR as compared to the pure H$_{2}$O case. 

Our results provide important implications for observing hot molten and young planets having short MO phases, especially for planets with a low initial volatile inventory and with an evolving atmosphere. The presence of H$_{2}$O, H$_2$, and perhaps O$_2$ in early thermal spectra of the planet could suggest about its evolutionary phase and also act as biomarkers for characterizing its atmosphere. 
%Furthermore, future observations of transmission spectra of planets such as GJ1132b and TRAPPIST-1c,1d could estimate the amount of H$_{2}$O etc. in the atmosphere. This could be possible in future for a range of hot magma ocean exoplanets.

Moreover, by including CO$_{2}$ as an another constituent of the atmosphere, and studying its thermal dissociation, this work invokes further investigation related to chemistry in the upper atmosphere leading to the possible formation of organics in the early Earth's atmosphere.

\section{Acknowledgements}
We thank the referee for useful comments that led to the improvement of the manuscript. NK acknowledges funding from the German Transregio Collaborative Research Centre ``Late Accretion onto terrestrial Planets (LATP)" [TRR170, sub-project C5]. MG acknowledges funding from DFG project (GO 2610/1-1). AN and NT acknowledges funding from the Helmholtz association (Project VH-NG-1017). FS acknowledge DFG project SCHR 1125/3-1. NT also acknowledges support from the German Research Foundation (DFG) through the SPP 1833 ``Building a habitable Earth" (project TO 704/2-1). NK thanks R. Wordsworth and L. Schaefer for providing OLR data for comparison in Fig.~\ref{f2} and Table 2. NK thanks S. St{\"a}dt for discussions on observer's geometry setup for GARLIC.

\newpage

\appendix

%\section{OLR sensitivity test based on $\beta$, hObs, and viewing angle $\alpha$.}
%Viewing angle is chosen to be $\alpha=$0$^{\circ}$,38$^{\circ}$. Below results are presented only for 38$^{\circ}$. The correct value of OLR is the one for hObs $>$ $R_{E}$; where radius of the Earth $R_{E}=6371$ Km. 

%\begin{figure}
%\plotone{figures/sens.pdf}
%\caption{Fig.5 \label{f5}}
%\end{figure}

\section{Comparison of OLR obtained using database HITRAN12 vs HITEMP10.}
\begin{figure*}[!hbt]
\gridline{\fig{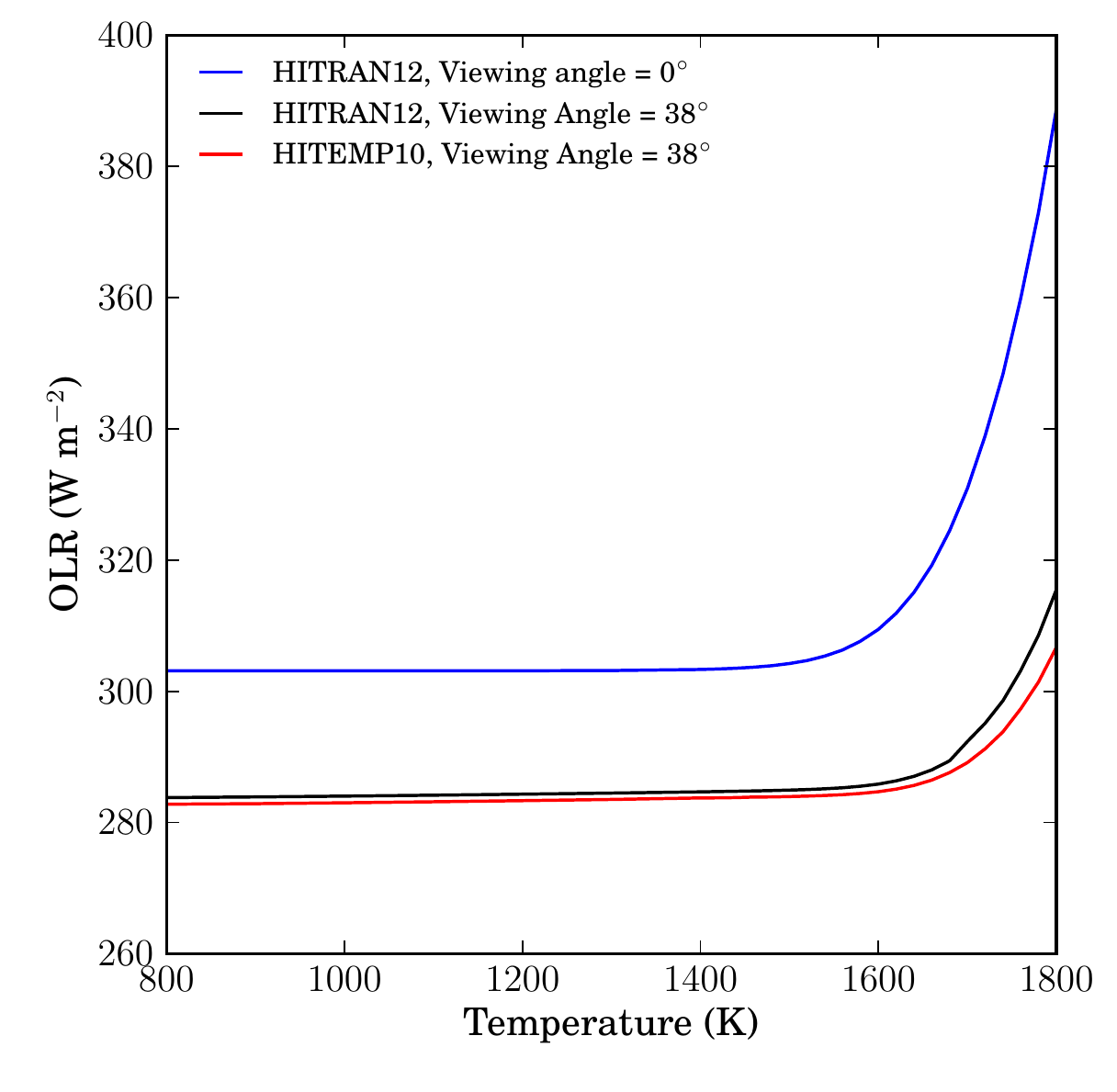}{0.3\textwidth}{(a)}
          \fig{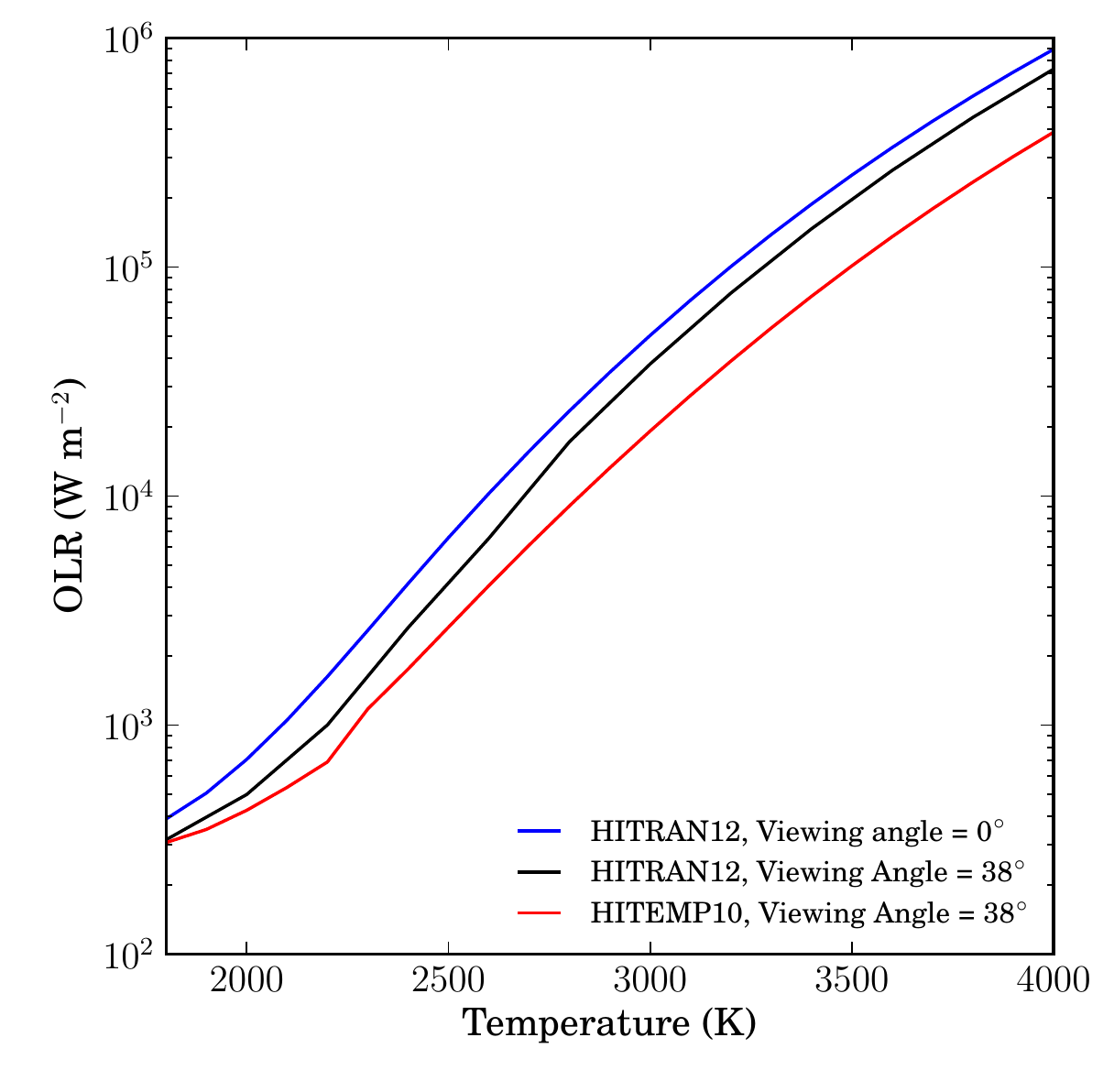}{0.3\textwidth}{(b)}
          \fig{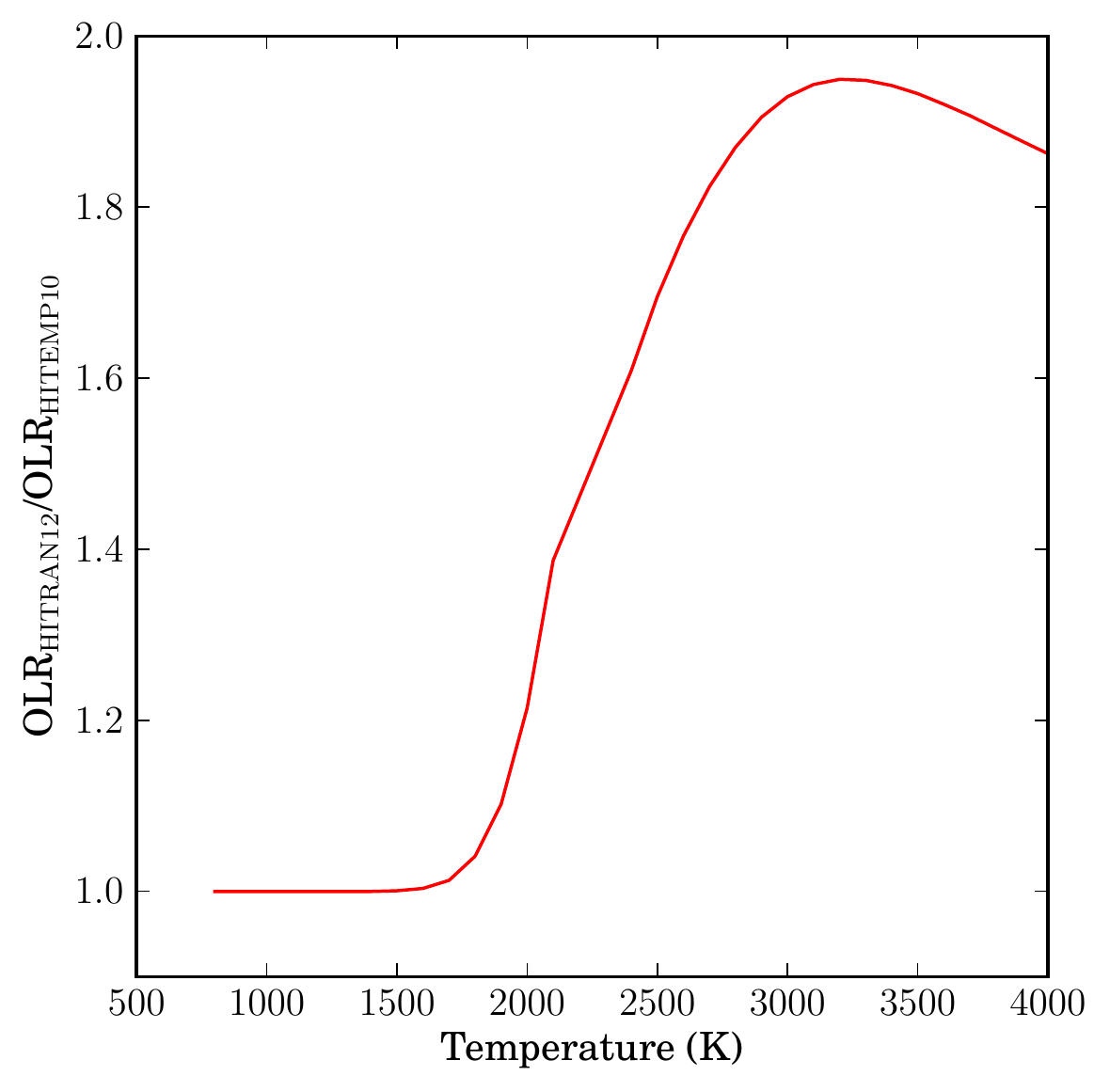}{0.3\textwidth}{(c)}
          }
%\gridline{\fig{RS_Oph.eps}{0.3\textwidth}{(d)}
%          \fig{U_Sco.eps}{0.3\textwidth}{(e)}
%          }
%\gridline{\fig{KT_Eri.eps}{0.3\textwidth}{(f)}}
\caption{(a) Comparison of OLR with different database HITRAN 2012 and HITEMP 2010; with different viewing angles for low temperature range $800 < T_{\text s} < \rm\, 1800 K$. (b) Same as (a) but for high temperature range $1800 < T_{\text s} < \rm\, \rm\, 4000 K$ and y-axis in log scale. (c) This panel shows the ratio of OLR calculated using HITEMP10 and HITRAN12 databases.  \label{f15}}
\end{figure*}

In this section, we compare the OLR obtained using spectral databases HITRAN12 \citep{rothman13} and HITEMP10 \citep{rothman10}. We also choose two different viewing angles: 0$^{\circ}$ and 38$^{\circ}$ and show the comparison of OLR vs. temperature in Fig.~\ref{f15}. As it can be seen, OLR is the same for the two databases up to $T_{\text s}= 1800 \rm\, K$ (Fig.~\ref{f15}a) and different for higher temperatures (Fig.~\ref{f12}b). We note that a lower OLR is obtained with HITEMP 2010 database as it includes more water lines at higher temperatures. The magnitude of the difference is shown in Fig.~\ref{f15}(c), i.e., we find that the OLR difference between the two databases rises monotonically after $T_{\text s}= 1800 \rm\, K$ and the OLR for the  HITRAN12 database is almost twice higher that HITEMP10 database around $T_{\text s}=3000 \rm\, K$.

\section{Comparison of effective height obtained using database HITRAN16 vs HITEMP10.}

\begin{figure*}[!hbt]
\centering
\includegraphics[scale=1]{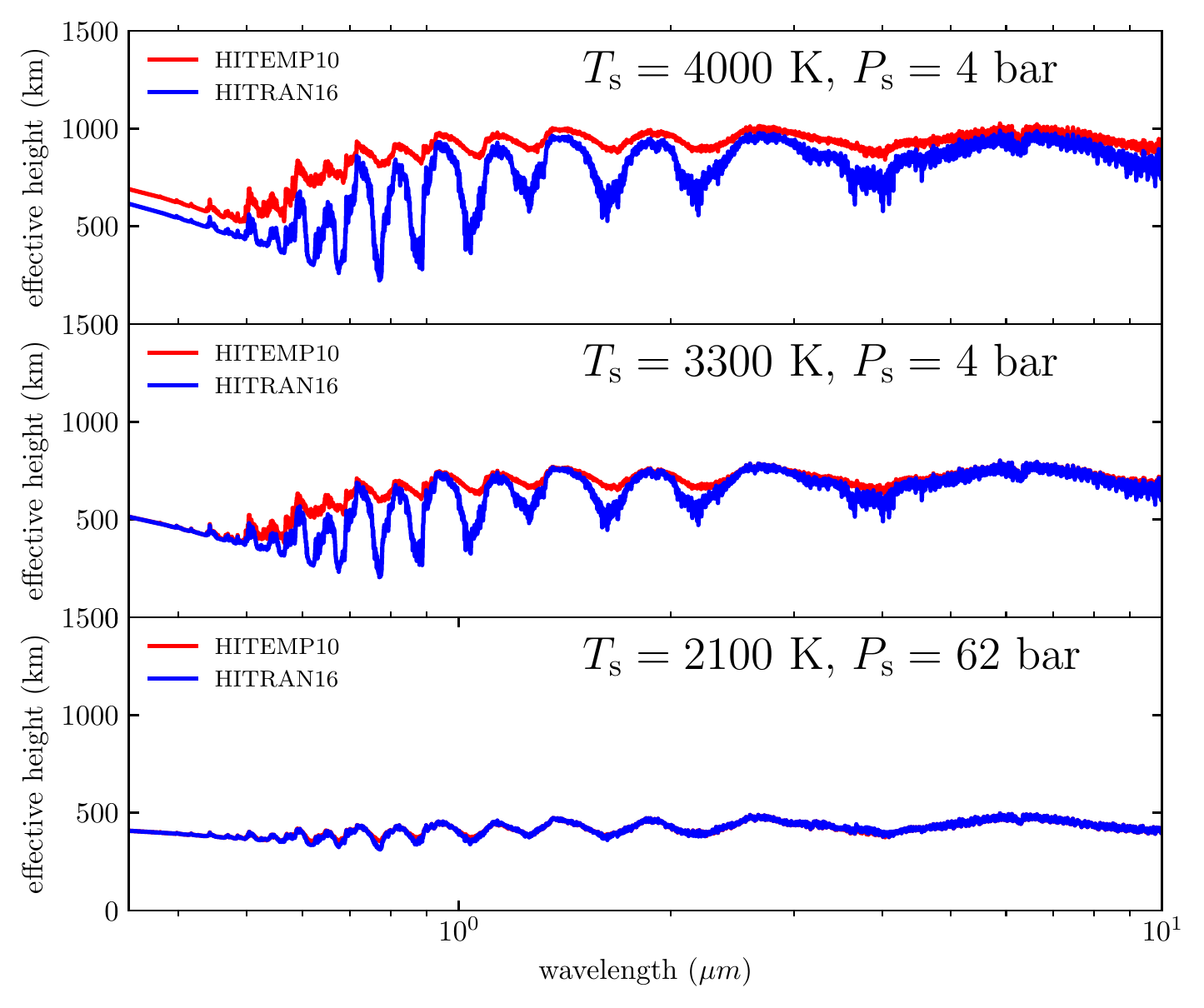}
\caption{Comparison of effective height of the atmosphere for pure water using two different databases.  \label{f16}}
\end{figure*}

\end{document}